\documentclass[onecolumn,
preprintnumbers,
nofootinbib,
amsmath,amssymb,
aps,
prd,
]{revtex4-2}

\usepackage{graphicx}
\usepackage{array}
\usepackage{dcolumn}
\usepackage{bm}% bold math
\usepackage{hyperref}
\usepackage{multirow}
\usepackage{slashed}
\DeclareUnicodeCharacter{2212}{\textendash}
\def\be{\begin{equation}}
\def\ee{\end{equation}}
\def\bea{\begin{eqnarray}}
\def\eea{\end{eqnarray}}

\usepackage{makecell}

\renewcommand{\d}{\delta}

\newcommand{\nn}{\nonumber}

\usepackage{xcolor}

\makeatletter
\newcommand{\Biggg}{\bBigg@{3.5}}
\makeatother

\begin{document}

\begin{flushright}
TUM-HEP-1568/25\\
NORDITA 2025-043\\
\vspace*{1cm}
\end{flushright}

\title{Hot New Early Dark Energy:\\ Dark Radiation Matter Decoupling}

\author{Mathias Garny$^1$}\email{mathias.garny@tum.de}
\author{Florian Niedermann${}^2$}\email{florian.niedermann@su.se}
\author{Henrique Rubira$^{3,4,5}$}\email{henrique.rubira@lmu.de}
\author{Martin S. Sloth${}^6$}\email{sloth@sdu.dk}
\affiliation{%
\small ${}^1$Physik Department T31, School of Natural Sciences, Technische Universit\"at M\"unchen \\
James-Franck-Stra\ss e 1, D-85748 Garching, Germany\\
${}^2$Nordita, KTH Royal Institute of Technology and Stockholm University\\
Hannes Alfv\'ens v\"ag 12, SE-106 91 Stockholm, Sweden
\\
${}^3$University Observatory, Faculty of Physics, Ludwig-Maximilians-Universit\"at, Scheinerstr. 1, D-81679 München, Germany\\
${}^4$Kavli Institute for Cosmology Cambridge, Madingley Road, Cambridge CB3 0HA, UK\\
${}^5$Centre for Theoretical Cosmology, Department of Applied Mathematics and Theoretical Physics University of Cambridge, Wilberforce Road, Cambridge, CB3 0WA, UK\\
${}^6$ Universe-Origins, University of Southern Denmark, Campusvej 55, 5230 Odense M, Denmark}%

\begin{abstract}
We present a microscopic model of the dark sector that resolves the Hubble tension within standard current data sets (Planck 2018, Pantheon+ and DESI DR2 BAO) based on well-known fundamental principles, gauge symmetry and spontaneous symmetry breaking. It builds on the Hot New Early Dark Energy (Hot NEDE) setup, featuring a dark $SU(N)$ gauge symmetry broken to $SU(N-1)$ in a supercooled phase transition that creates a thermal bath of self-interacting dark radiation in the epoch between Big Bang Nucleosynthesis and recombination. Adding a fermion multiplet charged under the gauge symmetry provides a naturally stable component of dark matter that interacts with dark radiation. Spontaneous symmetry breaking predicts a decoupling of this interaction once the dark sector cools down, that we refer to as dark radiation matter decoupling (DRMD). We also provide a simplified DRMD model that captures the essential features of the full theory while retaining additional falsifiable predictions. Using the data sets stated above, we find agreement with the SH${}_0$ES determination of $H_0$ at the 1.4$\sigma$ level, compared to a 5.7$\sigma$ tension in $\Lambda$CDM, thereby providing a resolution of the Hubble tension.

\end{abstract}

\maketitle
\begin{flushright}
{\it{``Such’ Er den redlichen Gewinn!\\
Sei Er kein schellenlauter Tor!\\
Es trägt Verstand und rechter Sinn\\
Mit wenig Kunst sich selber vor;\\
Und wenn’s euch ernst ist was zu sagen,\\
Ist’s nötig Worten nachzujagen?\\
Ja, eure Reden, die so blinkend sind,\\
In denen ihr der Menschheit Schnitzel kräuselt,\\
Sind unerquicklich wie der Nebelwind,\\
Der herbstlich durch die dürren Blätter säuselt!"}}

--- Johann Wolfgang von Goethe
\end{flushright}

\clearpage

\tableofcontents

\section{Introduction}
One of the main goals of particle cosmology is to use cosmological observations to probe the microphysics of the dark sector, supporting the discovery of new forces and forms of matter. In this work we argue that the observations leading to the Hubble tension allow us to approach this goal, pointing to dark sector physics governed by fundamental principles known from the visible sector to exist in Nature.

Local measurements of the expansion rate of the Universe are becoming increasingly accurate, and several different methods yield values inconsistent with those inferred indirectly from the Cosmic Microwave Background (CMB) and Baryon Acoustic Oscillation (BAO) measurements, when assuming the current standard model of cosmology, the $\Lambda$ Cold Dark Matter ($\Lambda$CDM) model. Most notably, the SH$_0$ES measurement of local type Ia supernovae (SNe) calibrated with Cepheids gives $H_0= 73.04\pm 1.04$ km/s/Mpc \cite{Riess:2021jrx} (and $H_0= 73.17\pm 0.86$ km/s/Mpc after a recent update~\cite{Breuval_2024}),  while a distance-ladder-free measurement using type II SNe finds $H_0= 74.9\pm 1.9$  (stat.\ only) km/s/Mpc~\cite{Vogl:2024bum}. In addition, there are measurements using type Ia SNe calibrated via the tip of the red giant branch (TRGB) instead of Cepheids yielding $H_0= 73.22\pm 2.06$  km/s/Mpc \cite{Scolnic:2023mrv}, or $70.4\pm2.1$ km/s/Mpc in the analysis of~\cite{Freedman:2024eph}. Other independent measurements are strong lensing time-delay measurements by the TDCOSMO collaboration obtaining $H_0=74.8\pm 3.5$ km/s/Mpc  when combining with BAO data from the DESI galaxy survey~\cite{Birrer:2025qbh}, and measurements using surface brightness fluctuations, yielding $H_0 = 73.3\pm 2.5$ km/s/Mpc~\cite{Blakeslee:2021rqi}, consistent with the other local measurements, but currently with larger error bars. 

In sharp contrast with the local measurements, the predicted value of $H_0$ within the $\Lambda$CDM model is $H_0= 67.36 \pm 0.54$ km/s/Mpc when using Planck 2018 CMB data \cite{Planck:2018vyg}. Therefore, unless measurements are systematically wrong across different methods and systematics, the $\Lambda$CDM model is disfavored with very high statistical significance. In the more extreme case, using SH$_0$ES data, the $\Lambda$CDM model is ruled out with more than $5\sigma$ significance (for a review on the Hubble tension and its proposed solutions see~\cite{CosmoVerse:2025txj}). Moreover, as more and more new data are being added, the Hubble tension tends to become more significant. For example, in their recent data release of ground-based CMB data, the South Pole Telescope (SPT) collaboration reports a $6.2 \sigma$ tension based on SPT data alone~\cite{SPT-3G:2025bzu}, with a similar trend seen by the Atacama Cosmology Telescope (ACT)~\cite{ACT:2025tim}. 
This prompts us to search for a new standard model of cosmology consistent with the data.

BAO measurements by the DESI collaboration provide an almost model-independent determination of the product $r_d \cdot H_0 = 101.54 \pm 0.73$ (100 km)/s \cite{DESI:2025zgx}, where  $r_d$ is the sound horizon at the time of baryon decoupling. This means that models fitting CMB data with a larger value of $H_0$ than $\Lambda$CDM must do so with also a smaller value of $r_d$ in order to keep the product  $r_d \cdot H_0$ in agreement with BAO measurements. Since the sound horizon $r_d$ only depends on the evolution of the Universe until just around recombination, models attempting to achieve agreement between CMB and the value of $H_0$ extracted from local measurements with only post-recombination physics generically fail badly in being consistent with BAO data. Thus, any solution to the Hubble tension that brings consistency between CMB data, $r_d \cdot H_0$ from BAO, and $H_0$ from local measurements must conservatively do so by modifying pre-recombination physics or the recombination history itself~\cite{Bernal:2016gxb,Knox:2019rjx,Benevento:2020fev}.

\medskip

 A simple and natural framework to address the Hubble tension by modifying pre-recombination physics is the New Early Dark Energy (NEDE) scenario \cite{Niedermann:2019olb,Niedermann:2020dwg,Niedermann:2021ijp,Niedermann:2021vgd,Cruz:2023lmn,Garny:2024ums}. In this setup the dark sector undergoes a fast-triggered phase transition (for recent reviews see \cite{Niedermann:2023ssr} or \cite{CosmoVerse:2025txj}, and for related work~\cite{Schoneberg:2021qvd,Freese:2021rjq,Allali:2021azp,Guendelman:2025swp}). Within the two most studied NEDE models, the Cold and the Hot NEDE models, a strong first-order phase transition is triggered either by an ultra-light axion in vacuum in the case of Cold NEDE \cite{Niedermann:2019olb,Niedermann:2020dwg,Cruz:2023lmn}, or by a non-vanishing dark sector temperature in Hot NEDE \cite{Niedermann:2021ijp,Niedermann:2021vgd,Garny:2024ums}.  In both cases, the vacuum energy in the false vacuum, i.e.\ the latent heat associated with the phase transition, becomes important in the radiation-dominated Universe before the phase transition, and is converted into a dark fluid after the phase transition. This leads to an injection of extra energy, stemming from the latent heat and/or the post-phase-transition fluid, which reduces the sound horizon and allows for a solution to the Hubble tension.
 
 The Cold and Hot NEDE models, however, differ in important testable ways. In the Cold NEDE model the vacuum energy leads to an energy injection right before matter domination, which quickly decays away in a stiff NEDE fluid with equation of state $w\sim 2/3$. Since the energy density of this fluid decays away fast, the Cold NEDE phase transition has to happen relatively close to recombination, like in the axiEDE model \cite{Poulin:2018cxd} (see \cite{Poulin:2023lkg} for a review), although in axiEDE the transition happens even later between matter-radiation equality and recombination. Moreover, these models are realized by completely different microphysics, with the axiEDE model invoking an arguably exotic axion potential~\cite{Kaloper:2019lpl}. The Cold NEDE and axiEDE models are therefore distinct models both conceptually and also phenomenologically.
 
 In the Hot NEDE scenario, which is the focus of this work, the latent heat of the supercooled phase transition is instead converted into dark radiation, which is not diluted away compared to ordinary radiation. It is the extra dark radiation that then leads to a decrease of the sound horizon, and the phase transition can therefore happen anywhere between Big Bang Nucleosynthesis (BBN) and recombination. In order to have a large enough effect on the sound horizon, it is important that the extra dark radiation is created {\it after} BBN, because otherwise it would spoil the successful prediction of light element abundances. 
 In this respect, Hot NEDE UV-completes models of Strongly Interacting Dark Radiation (SIDR)~\cite{Jeong:2013eza,Buen-Abad:2015ova,Buen-Abad:2017gxg,Archidiacono:2020yey,Blinov:2020hmc,Aloni:2021eaq}, as it provides a way to create the large value of $\Delta N_\text{eff}$ assumed in SIDR models while being consistent with BBN~\cite{Garny:2024ums}. Furthermore, the Hot NEDE model does lead to a number of new testable predictions, like a sharp feature in the matter power spectrum at small scales and a gravitational wave signal, which can be searched for using pulsar timing arrays~ \cite{Garny:2024ums}. Finally, while SIDR models cannot fully resolve the Hubble tension, we find such a solution in this work resulting from the specific microphysics of Hot NEDE.

\begin{figure}
    \centering
    \includegraphics[width=0.99\linewidth]{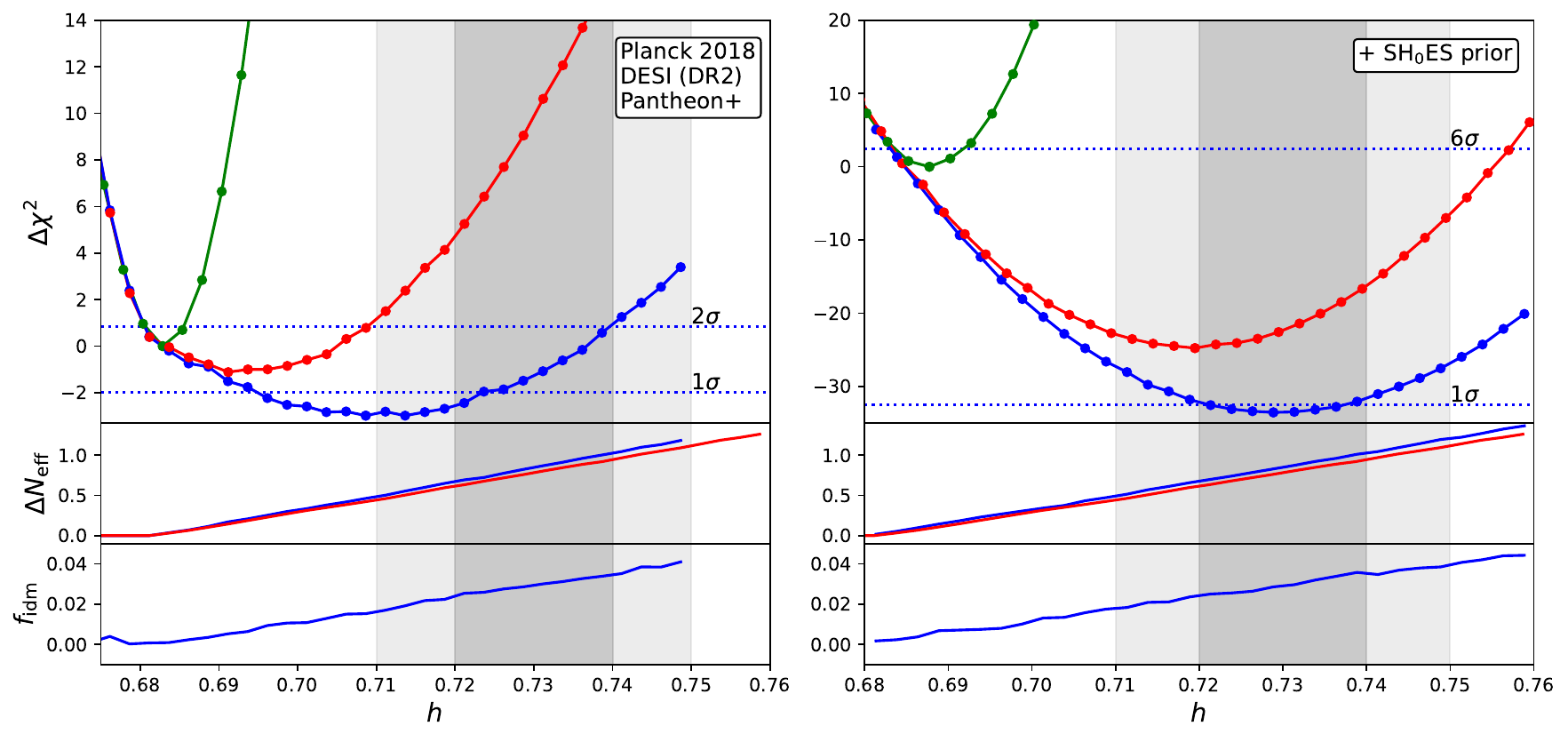}

    \includegraphics[width=0.5\linewidth]{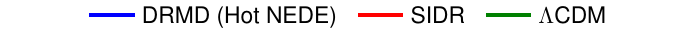}

    \caption{Profile likelihood curves for the DRMD dark sector model, arising as a particular limit of the Hot NEDE model and proposed in this work (blue), compared to a model of self-interacting dark radiation (SIDR, red) and to $\Lambda$CDM (green). The left panel includes Planck 2018 CMB, DESI BAO (DR2), and (uncalibrated) Pantheon+ SNe data, while the right panel adds in addition the local $H_0$ measurement from SH$_0$ES in the likelihood (the $1$- and $2\sigma$ regions preferred by SH$_0$ES are also indicated by the gray regions in both panels for comparison). The lower panels show the model parameters of (post-BBN) $\Delta N_\text{eff}$ and $f_\text{idm}$ (fraction of interacting dark matter). The feature of decoupling between dark radiation and dark matter within the DRMD model impacts the evolution of perturbations, making it consistent with large values of $h=H_0 / (100\,\mathrm{km\,s^{-1}\,Mpc^{-1}})$ and $\Delta N_\text{eff}$. The dark radiation is produced in a phase transition after BBN, while $\Delta N_\text{eff}\ll 1$ during BBN. 
    The low-$h$ regime corresponds to vanishing $\Delta N_{\rm eff}$, and the three models coincide.
    The DRMD model follows from a well-defined microscopic theory based on known principles, including gauge symmetry and spontaneous symmetry breaking in the dark sector.  
    }
    \label{fig:profile}
\end{figure}

\medskip

We consider a microphysical implementation of the Hot NEDE model guided by well-known fundamental principles from the perspective of particle physics, encompassing gauge symmetries and spontaneous symmetry breaking. We assume a dark sector governed by an $SU(N)$ gauge symmetry, that is spontaneously broken to $SU(N-1)$ via the Higgs mechanism, following~\cite{Garny:2024ums}. The associated phase transition features supercooling~\cite{Witten:1980ez} if symmetry breaking occurs via the Coleman-Weinberg mechanism~\cite{Coleman:1973jx}, and creates a thermal plasma of dark radiation with sizeable $\Delta N_\text{eff}$ after BBN. Furthermore, non-Abelian gauge boson self-interactions ensure that the dark radiation behaves like a fluid and does not free-stream~\cite{Buen-Abad:2015ova,Lesgourgues:2015wza,Buen-Abad:2017gxg,Rubira:2022xhb}.

In this work, we extend the Hot NEDE model by including a massive Dirac fermion multiplet transforming in the fundamental representation of the $SU(N)$ gauge group of the dark sector, while being a singlet under the Standard Model gauge symmetries. Gauge and Lorentz symmetries make it stable, and thus a candidate to constitute part of dark matter. Its coupling to gauge bosons leads to an interaction between dark matter and dark radiation via a drag force~\cite{Buen-Abad:2015ova,Lesgourgues:2015wza,Buen-Abad:2017gxg,Rubira:2022xhb} (see also~\cite{vandenAarssen:2012vpm,Cyr-Racine:2015ihg,Chacko:2016kgg,Binder:2016pnr}), and initially tightly couples both fluids. Spontaneous symmetry breaking induces a small mass splitting among the components of the multiplet, with the lighter one being neutral under the residual $SU(N-1)$ symmetry. An analogous property is well-known in the context of electroweak symmetry breaking for supersymmetric dark matter (see e.g.~\cite{Ibe:2012sx}). However, here it appears within the dark sector, leading to a very different phenomenology. Specifically, once the dark sector temperature drops below the mass splitting, the interactions turn off in a process we refer to as {\it Dark Radiation Matter Decoupling} (DRMD). Importantly, the dark radiation remains fluid-like even after this decoupling, since the self-interactions among the massless $SU(N-1)$ gauge bosons remain active.
We dub this setup the DRMD model. 
We work out and implement the evolution of perturbations in the DRMD model into a modified version of \texttt{CLASS}~\cite{Blas:2011rf}, and test it against Planck 2018 legacy data alongside the recently released DESI BAO (DR2) and Pantheon+ SNe data~\cite{Brout:2022vxf}.

Let us briefly preview our main results. Within our baseline data combination, consisting of Planck 2018 CMB, BAO and (uncalibrated) SNe data, we find consistency with the local determination of $H_0$ by SH$_0$ES at the $1.4\sigma$ level with (effectively) only three\footnote{The MCMC analysis samples four parameters. However, one of the results of this work is that two of these parameters are highly degenerate, implying that one of them can be fixed to an arbitrary value within the prior range without affecting our conclusions. While future analyses may therefore adopt the corresponding three-parameter description, we chose not to repeat the present analysis in this trivially reduced parameter space.}
extra free parameters compared to $\Lambda$CDM. In this sense, the model provides a full resolution of the Hubble tension for this baseline data set.
The specific DRMD dynamics has an impact on the evolution of perturbations such that CMB and BAO data allow for a relatively large post-BBN value of $\Delta N_\text{eff}$, provided that the decoupling between dark radiation and dark matter occurs at a redshift $z_\text{dec}$ around matter-radiation equality, and that only a percent-level fraction $f_\text{idm}$ of dark matter interacts, while the remaining part behaves as CDM (being e.g.\ provided by an additional field that is a singlet under $SU(N)$).
Since our Bayesian analysis indicates the presence of projection effects, we also perform a profile likelihood analysis, finding consistent results regarding the ability of the model to resolve the Hubble tension (see \cite{Herold:2021ksg} for a discussion of profile likelihood analysis in the context of EDE or \cite{Cruz:2023cxy} for Cold NEDE). 

We show the profile likelihood for the DRMD model studied in this work with respect to $H_0$ in Fig.\,\ref{fig:profile}, for the case when including Planck 2018 CMB, DESI BAO (DR2), and (uncalibrated) Pantheon+ SNe  data (left panel), and when combining also with the local $H_0$ measurement by SH$_0$ES implemented through a Gaussian prior on the absolute SNe magnitude (right panel). For comparison, we also show the profile likelihoods for the standard $\Lambda$CDM model as well as an SIDR model (with fluid-like extra relativistic energy, assumed to be created after BBN, but no interaction with dark matter).
Even when not including any local $H_0$ measurements in the left panel, the DRMD model prefers large values in a wide range $h\simeq 0.69-0.73$, featuring a flat $\Delta\chi^2$ such that $h=0.73$ is allowed at the $\sim 1.5\sigma$ level, in stark contrast to the SIDR model. The right panel shows the resulting significant improvement $\Delta\chi^2\simeq -33$ when including the local SH$_0$ES measurement for DRMD. The lower panels depict the model parameters $\Delta N_\text{eff}$ and $f_\text{idm}$. 

We note that the specific aspect of an interaction rate between dark radiation and dark matter components that becomes suppressed at some point during cosmological evolution has been considered in completely different microphysical setups, such as dark recombination in atomic dark matter~\cite{Kaplan:2009de,Cyr-Racine:2012tfp,Cyr-Racine:2021oal,Blinov:2021mdk,Bansal:2022qbi,Greene:2024qis} or due to the mass of the particle mediating the interaction~\cite{vandenAarssen:2012vpm,Cyr-Racine:2015ihg,Joseph:2022jsf}. Even more closely related phenomenology arises also in stepped partially acoustic (SPartAcous) models, in which a subcomponent of dark matter is coupled to dark radiation through a drag force whose efficiency decreases exponentially during cosmological evolution when a massive dark radiation species becomes non-relativistic, creates an extra `step' (which actually makes SPartAcous a phenomenologically less viable as solution to the Hubble tension), and annihilates~\cite{Chacko:2016kgg,Buen-Abad:2022kgf,Buen-Abad:2023uva} (see also \cite{Allali:2023zbi,Schoneberg:2023rnx}). Interactions between dark matter and dark radiation described by a drag force have also been discussed in e.g.~\cite{vandenAarssen:2012vpm,Lesgourgues:2015wza,Cyr-Racine:2015ihg,Vogelsberger:2015gpr,Buen-Abad:2015ova,Buen-Abad:2017gxg,Chacko:2016kgg,Rubira:2022xhb, Hooper:2022byl,Mazoun:2023kid, SPT:2024roj, Euclid:2024pwi}. 
However, these setups typically differ in their cosmological evolution from the Hot NEDE model, as well as in the details of how the interaction becomes suppressed, or whether the dark radiation remains fluid-like or starts to free-stream, which turn out to be relevant for the ability to address the Hubble tension. The novelty of the present construction is therefore not the existence of a DM--DR drag force by itself, but rather its microscopic origin within the Hot NEDE setup. The same symmetry-breaking event that allows the dark radiation abundance to be generated after BBN also induces a loop-level mass splitting in the dark matter multiplet, leading to an exponential suppression of the interacting charged states.

\medskip

This work is structured as follows: In Sec.~\ref{sec:micro}, we first provide an overview of the microscopic model and its cosmological evolution, with a subsequent detailed discussion including a review of the phase transition dynamics following~\cite{Garny:2024ums} as well as the novel aspects related to the dark matter multiplet. The evolution of perturbations  is discussed in Sec.~\ref{sec:pheno}, first for the full model and then for a simplified setup capturing only the essential DRMD features, that we employ in our data analysis. In Sec.~\ref{sec:results} we present our results and conclude in Sec.~\ref{sec:conclusion}. The appendixes contain a table with further information on our results (Appendix~\ref{sec:App_table}) as well as computational details.

\section{Microscopic model}
\label{sec:micro}

In this section we discuss the microphysical model. We assume a dark sector that is sequestered from the visible sector, but governed by known fundamental principles, i.e.~gauge symmetry complemented by the Higgs mechanism. Taking these as inspiration, we consider a dark $SU(N)$ gauge symmetry, that is spontaneously broken to $SU(N-1)$ by a dark Higgs field $\Psi$~\cite{Garny:2024ums}. Moreover, in this work, we also incorporate a fermionic multiplet $\chi$ charged under $SU(N)$ that describes (part of) dark matter (while the remaining part is assumed to be standard CDM). Before a more detailed discussion, we provide an overview of the main features and the resulting cosmological evolution.

\subsection{Overview}\label{sec:overview}

\begin{figure}[!t]
  \begin{center}
  \includegraphics[width=0.6\textwidth]{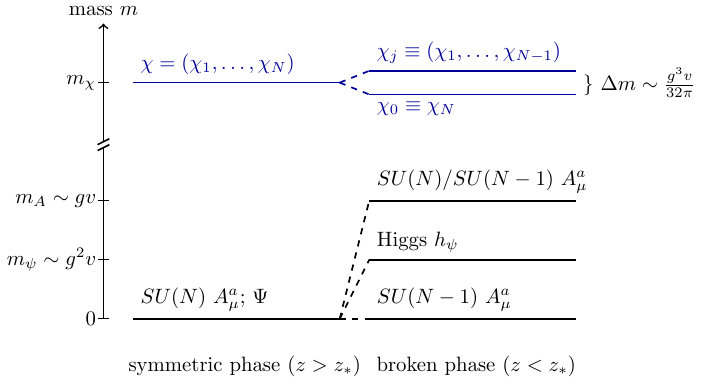}
  \end{center}
  \caption{Mass spectrum of the dark sector in the unbroken $SU(N)$ phase ($z>z_*$) and in the broken $SU(N-1)$ phase ($z<z_*$). The Higgs mechanism generates masses $m_A\sim gv$ of all gauge bosons $A_\mu^a$ except for those belonging to the residual $SU(N-1)$ subgroup, forming a component of self-interacting dark radiation (SIDR). The dark Higgs has a mass $m_\psi\sim g^2v$, being characteristic for the Coleman-Weinberg mechanism. In blue, the fermion multiplet $\chi$ is shown, being (part of) dark matter. It receives a loop-induced mass splitting $\Delta m$, which makes only the neutral component $\chi_0$ stable, while the charged $\chi_j$ states convert to the neutral one at temperatures $T_d\ll \Delta m$.}
  \label{fig:spectrum}
\end{figure}

The Lagrangian describing the microphysics of the dark sector is largely dictated by the assumed gauge symmetry, and given by the canonical form
\be
{\cal L} = \bar\chi (i\gamma^\mu D_\mu-m_\chi)\chi + |D\Psi|^2 - V_{\rm cl}(|\Psi|^2) -\frac14 F^a_{\mu\nu}F_a^{\mu\nu} \,,
\ee
with $F_a^{\mu\nu}$ being the field strength tensor of the dark gauge fields $A_a^\mu$, associated with the generators $\tau_a$ of $SU(N)$. We assume both the dark Higgs field $\Psi=(\Psi_1,\dots,\Psi_N)$ and the dark matter multiplet $\chi=(\chi_1,\dots,\chi_N)$ to transform under the fundamental representation of $SU(N)$, such that their gauge interaction is determined by the usual covariant derivative $D^\mu  = \partial^\mu  - ig A_a^\mu \tau^a $ with gauge coupling $g$ assumed to be perturbatively small, i.e.~$g^2/(4\pi)\ll 1$. The potential $V_{\rm cl}$ is given by the standard form, see~\eqref{eq:vac_pot} below, and gives rise to spontaneous symmetry breaking $SU(N)\to SU(N-1)$ via the Higgs mechanism, with vacuum expectation value (VEV) that can without loss of generality be parametrized as $\langle\Psi\rangle=(0,\dots,0,v/\sqrt{2})$. For simplicity we assume $\chi$ to be a Dirac fermion, allowing for a Dirac mass term but precluding a Yukawa interaction with $\Psi$. The relevant fundamental parameters are thus the VEV $v$, the gauge coupling $g$ and the dark matter mass scale $m_\chi$. Following~\cite{Garny:2024ums} we assume the Higgs potential to be (close to) classically scale-invariant, such that $v$ is generated by dimensional transmutation via the Coleman-Weinberg mechanism~\cite{Coleman:1973jx} (being parametrically the scale at which the running quartic Higgs coupling $\lambda(v)$ is of the order of $g^4$, thus trading $\lambda$ for $v$ as model parameter). The dimensionful scales are therefore `protected' by chiral and scale symmetries, respectively. 

The mass spectrum in the symmetric and the broken phase is illustrated in Fig.~\ref{fig:spectrum}. In the latter, the gauge bosons related to the residual $SU(N-1)$ subgroup remain massless, while the other ones acquire a mass of order $m_A\sim gv$ and famously absorb all (Goldstone) components of the Higgs multiplet $\Psi$ except for a single degree of freedom, being the (dark) Higgs boson. The Coleman-Weinberg mechanism requires its mass to be of order $m_\psi\sim g^2v$. Finally, $\chi$ splits into the two subsets $\chi_0\equiv\chi_N$ and $\chi_j\equiv(\chi_1,\dots,\chi_{N-1})$, being neutral and charged under the residual $SU(N-1)$, respectively. Radiative corrections induce a (small) mass splitting $\Delta m\sim g^3v/(32\pi)$, making $\chi_0$ slightly lighter. Notably, the splitting $\Delta m$ is independent of $m_\chi$ for $m_\chi\gg m_A$~\cite{Ibe:2012sx}, which we assume in the following. Apart from this, the ordering of mass scales generated via spontaneous symmetry breaking, 
\be \label{eq:mass_hierarchy}
  m_A\sim gv\gg m_\psi\sim g^2v\gg \Delta m\sim g^3v/(32\pi)\,,
\ee
is a direct consequence of the microphysical setup (with viable values of the gauge coupling being $g\sim {\cal O}(0.1)$). This mass hierarchy has a decisive impact on its cosmological evolution, with specific dynamics linked to the redshifts at which the temperature of the dark sector drops below each of those mass scales.

\medskip

\begin{figure}[!t]
  \begin{center}
  \includegraphics[width=0.65\textwidth]{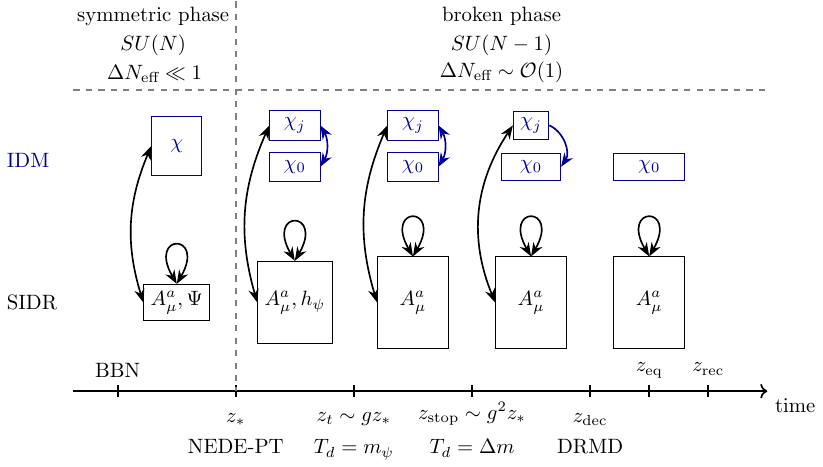}
  \end{center}
  \caption{Schematic illustration of the cosmological evolution of the dark sector, consisting of self-interacting dark radiation (SIDR) and interacting dark matter (IDM). During BBN the dark sector contribution to the total radiation energy is negligible. The vacuum energy released in the supercooled NEDE phase transition (NEDE-PT) at $z=z_*$ reheats the dark sector, and leads to a splitting between charged ($\chi_j$) and neutral ($\chi_0$) IDM components. At $z=z_t$ the Higgs drops out of the SIDR plasma, and at $z=z_\text{stop}$ the occupation of $\chi_j$ becomes Boltzmann-suppressed relative to $\chi_0$. This leads to a termination of the tight coupling between IDM and SIDR at $z=z_\text{dec}$ (dark radiation matter decoupling, DRMD), resolving the Hubble tension if $z_\text{dec}$ occurs around matter-radiation equality ($z_\text{eq}$), slightly before recombination ($z_\text{rec}$).  For comparison with the stepped SIDR model, we refer to Fig.~1 of~\cite{Schoneberg:2023rnx}. Black arrows indicate energy/momentum exchange in elastic processes such as e.g.~$\chi_jA_\mu^a\leftrightarrow\chi_j A_\mu^a$, and blue arrows additionally conversions among $\chi_j$ and $\chi_0$ states. }
  \label{fig:evolution}
\end{figure}

An outline of the cosmological evolution, as discussed in detail below, is as follows:
\begin{itemize}
\item The (secluded) dark sector is initially populated in the Early Universe, and is then subsequently completely decoupled from the visible sector. While the precise mechanism is irrelevant here, a suitable possibility is freeze-in~\cite{Hall:2009bx} (e.g.\ minimally via gravitational interactions~\cite{Garny:2015sjg}). Dark gauge interactions establish thermal equilibrium in the dark sector, at some initial temperature $T_d\ll T_\text{vis}$ that we assume to be (much) smaller than the one in the visible sector\footnote{In the self-interacting Planckian interacting dark matter (PIDM) scenario one naturally has $T_d \sim 10^{-3} T_\textrm{vis}$ \cite{Garny:2018grs}.}.
\item Once $T_d$ drops below $m_\chi$ the abundance of $\chi$ remains fixed, and provides a constant fraction $f_\chi\equiv f_\text{idm}$ of the total dark matter density. The details of this dynamics are left unspecified in this work, and we treat $f_\chi$ as a free parameter. A generic scenario is freeze-out of $\chi$ within the dark sector into dark gauge and Higgs bosons, featuring an interesting but also rather complex dynamics~\cite{Harz:2018csl,Binder:2021otw,Binder:2023ckj}. Subsequently, the dark sector consists of a SIDR component formed by the gauge and Higgs bosons, and a non-relativistic interacting dark matter (IDM) component, given by the $\chi$ multiplet, that interacts with the SIDR via gauge interactions.
\item During the BBN epoch, $z\sim 10^9$, the low initial dark sector temperature leads to a negligible relativistic energy density, denoted by $\Delta N_\text{BBN}\ll 1$. Its precise value is irrelevant for the subsequent evolution, but ensures that standard BBN predictions are not spoiled by the dark sector.
\item At some redshift $z=z_*$, after the BBN epoch, the dark sector undergoes a first-order phase transition $SU(N)\to SU(N-1)$. As is well-known, the Coleman-Weinberg setup leads to strong supercooling, with a sizeable vacuum energy being released into the dark sector~\cite{Witten:1980ez}. The value of $z_*$ is set by the small deviation from scale-invariance in the Higgs potential, related to the quadratic term. The latter only affects $z_*$, being therefore effectively a free model parameter. The phase transition is assumed to be after BBN but sufficiently long before recombination, roughly $10^4\ll z_* \lesssim 10^9$.
\item The latent heat increases the entropy of the dark sector and reheats it to a temperature $T_d\sim gv$ after the phase transition, creating a sizeable contribution to the radiation energy, 
\be\label{eq:NeffNEDE}
  \Delta N_\text{eff} \equiv \Delta N_\text{NEDE}\sim \left(\frac{g\,v}{85\,\text{keV}}\right)^4\left(\frac{10^8}{1+z_*}\right)^4 \quad (\text{for}\,\,N=3)\,, 
\ee
carried by the massless $SU(N-1)$ gauge bosons and the Higgs boson~\cite{Garny:2024ums}. 
This energy injection is primarily responsible for lowering the sound horizon $r_d$, and the resulting order of magnitude $\Delta N_\text{NEDE}\sim {\cal O}(1)$ to address the Hubble tension  sets the overall phase transition scale $gv$. Since $m_\chi\gg gv$ we assume that the $\chi$ abundance is not affected by the phase transition, even though a more detailed study of this point would be interesting (see e.g.~\cite{Giudice:2024tcp}). The $\chi_0$ component becomes slightly lighter than $\chi_j$ due to the mass splitting $\Delta m$. However, since $T_d\sim gv\gg \Delta m$, all components of $\chi$ are equally populated at this time, as dictated by equipartition.
\item Gauge interactions ensure that conversions among the $\chi_0$ and $\chi_j$ components are in equilibrium. The charged states efficiently Compton scatter $\chi_j A^\mu_a\leftrightarrow \chi_j A^\mu_a$ off the massless gauge bosons, leading to a drag force that is enhanced due to the non-Abelian interaction~\cite{Moore:2004tg,Buen-Abad:2015ova,Cyr-Racine:2015ihg,Rubira:2022xhb} (as compared to the analogous baryon-photon scattering process). Overall, this creates a single, tightly coupled fluid in the dark sector with common sound velocity $c_s\ll 1/\sqrt{3}$.
\item At redshift $z=z_t\sim gz_*$ the temperature $T_d$ drops below $m_\psi\sim g^2v$ and the Higgs becomes non-relativistic. It drops out of the plasma and slightly increases $\Delta N_\text{eff}$ (while preserving entropy) due to the mass-threshold effect~\cite{Aloni:2021eaq}, leading to a slight further increase to the final value $\Delta N_\text{IR}$~\cite{Garny:2024ums}. While this `second step' is a prediction of the model, we find its phenomenological impact to be small. Moreover, the size of this second step vanishes in the large-$N$ limit of the dark $SU(N)$.
\item At $z=z_\text{stop}\sim g^2z_*$, the temperature $T_d$ becomes smaller than the mass splitting $\Delta m$. 
Subsequently the heavier charged components $\chi_j$ of the dark matter multiplet become Boltzmann-suppressed relative to $\chi_0$, $n_{\chi_j}/n_{\chi_0}\sim e^{-\Delta m/T_d}=e^{-z_\text{stop}/z}$. Nevertheless, the drag force due to gauge interaction still maintains tight coupling, but the effective rate quickly drops since the massless gauge bosons effectively Compton scatter only off the charged $\chi_j$ states. 
\item At $z=z_\text{dec}\sim z_\text{stop}/30$ tight coupling breaks down. Afterwards, the massless gauge bosons form a fluid component of self-interacting dark radiation (i.e.\ with $c_s=1/\sqrt{3}$ but no anisotropic stress), and the dark matter (now being predominantly in the neutral state $\chi_0$) behaves as distinct matter component approaching CDM behavior. This is the process of DRMD which we find is crucial to resolve the Hubble tension if $z_\text{dec}$ occurs somewhat before recombination. For the scales we consider, the ratio $z_\text{dec}/z_\text{stop}$ is the only place where $m_\chi$ enters, with only a logarithmic sensitivity, see~\eqref{eq:z_ratio}.
\end{itemize}

A sketch of the cosmological history is shown in Fig.~\ref{fig:evolution}. The essential DRMD dynamics is captured by the last two items, which are the focus of the effective DRMD model introduced in Sec.\,\ref{sec:DRMD}. It is applicable for CMB observables in the limit in which the phase transition ($z=z_*$) occurs after BBN but well before modes measured in the CMB angular power spectra enter the horizon, roughly $10^6\lesssim  z_* \lesssim 10^9$, and if the `second step' at $z=z_t$ is sufficiently small (large-$N$ limit of $SU(N)$). Nevertheless, we note that the full model, including the phase transition, is relevant for predicting features in the matter power spectrum at (small) scales entering the horizon at $\sim z_*$, for a potential gravitational wave signal either from the scalar field \cite{Huber:2008hg,Caprini:2007xq,Jinno:2016vai} or from fluid sound waves \cite{Hindmarsh:2013xza,Hindmarsh:2015qta,Jinno:2022mie,Caprini:2024gyk} which is especially compelling in light of the recent PTA results \cite{EPTA:2023fyk, NANOGrav:2023gor}, and for consistency with BBN~\cite{Garny:2024ums}. A further potential signature of the phase transition are traces of stochasticity from the finite size of bubbles when they percolate~\cite{Niedermann:2020dwg,Elor:2023xbz}. A naive comparison to constraints derived assuming this effect on top of $\Lambda$CDM perturbations suggests that they are easily satisfied for phase transitions that occur either sufficiently early ($z_*\gtrsim 10^6-10^7$) or on a time-scale (denoted conventionally by $1/\beta$) that is a sufficiently small fraction of the Hubble time $H_*$, roughly $\beta /H_*\gtrsim {\cal O}(10-100)$ for $z_*\lesssim 10^6$. Another possibility is that these small-scale anisotropies induce spectral $\mu$-distortions in the CMB which could in principle be probed in the future \cite{Ramberg:2022irf}.

\medskip

From a broader perspective, we note formal similarities of the DRMD mechanism to the breakdown of tight coupling among $e^-$, $p$ and photons in the visible sector due to recombination, in the sense that the abundance of electrically charged states (i.e. $e^-$ and ionized hydrogen) becomes small below a certain redshift, in analogy to the $\chi_j$ abundance at $z<z_\text{stop}$ as considered here. However, in stark contrast to the photons within the visible sector, the dark radiation does {\it not} start to free-stream after decoupling from the IDM component at $z>z_\text{dec}$, but rather remains described by a SIDR fluid due to its non-Abelian self-interactions (meaning specifically that its sound velocity increases due to DRMD, but anisotropic stress and higher multipoles remain small). Moreover, the microphysics and resulting processes and rates are very different. Nevertheless, keeping the formal analogy to certain aspects of recombination in mind, it would be interesting to investigate further details of the DRMD process. We refer to Sec.~\ref{sec:decoupling} for further discussion of this point.

Once more, we emphasize the conceptual simplicity of the model considered in this work, with a dark $SU(N)$ gauge symmetry and fermions coupled to this dark sector. The model contains four micro-physical parameters: $g, v$ (or alternatively $\lambda$), the effective Higgs mass $\mu_\mathrm{eff}$ and $m_\chi$, the last of which is effectively irrelevant, together with the discrete gauge group parameter $N$. These are complemented with $f_{\rm idm}$ and the redshift of the phase transition $z_*$ (or alternatively $\Delta N_{\rm eff}$).\footnote{Also related to the NEDE fraction in \cite{Garny:2024ums}.} As we describe later and as already mentioned, effectively only three parameters describe the  macrophysics of the DRMD fluid relevant to the Hubble tension, being $\Delta N_{\rm eff}$, $f_{\rm idm}$ and $z_\text{dec}$. We also refer to Table~I of \cite{Garny:2024ums} for the typical values of parameters related to the phase transition.

We note that in~\cite{Schoneberg:2023rnx} a model exhibiting a similar DRMD mechanism was considered (dubbed strongly interacting stepped radiation or SPartAcous, see also~\cite{Buen-Abad:2022kgf}). However, due to the different microphysics, in that scenario a sizeable `step' in $\Delta N_\text{eff}$ occurs coincidentally with an onset of Boltzmann suppression of the drag rate between IDM and SIDR. We identify this distinct feature as a cause for the finding from~\cite{Schoneberg:2023rnx} regarding the moderate success of that setup in addressing the Hubble tension, in contrast to the model discussed in this work, where $\Delta N_\text{eff}$ is constant during the DRMD dynamics.\footnote{Since the microscopic model considered in~\cite{Schoneberg:2023rnx} requires the presence of a step in $\Delta N_\text{eff}$ at the same redshift for which exponential suppression of IDM-DR interaction sets in, the authors did not investigate the limiting case for which such a step is absent in detail. Nevertheless, in Fig.~13 of~\cite{Schoneberg:2023rnx} a scenario with variable stepsize is analyzed, reporting a preference towards small values and improved~$\chi^2$. This setting also prefers a value of $\log_{10}(z_t)$, which via Fig.~2 appears to be compatible with a decoupling around matter-radiation equality. However, this feature is not discussed in their work. Moreover those results combine CMB with SH$_0$ES and also $S_8$ data, such that a direct comparison to our work is difficult. This also means that no quantification of the (potential relaxation of the) Hubble tension for small (or variable) step size is possible from the results shown in Fig.~13 of~\cite{Schoneberg:2023rnx}.}
Another scenario in which an IDM component stops interacting with a dark radiation component is `atomic dark matter'~\cite{Kaplan:2009de,Cyr-Racine:2012tfp,Cyr-Racine:2021oal,Blinov:2021mdk,Bansal:2022qbi}, featuring a dark recombination process.  However, in that setup the dark radiation starts to free-stream after dark recombination, similarly as for photons, making this scenario rather different from the one considered here at the level of perturbations, although in the `new atomic dark matter' model radiation after dark recombination is self-interacting leading to a better data fit~\cite{Buen-Abad:2024tlb} (when ignoring BBN constraints). In any event, we note that the mechanism of DRMD as predicted by the $SU(N)$ dark sector could still be realized in different microphysical setups than the one we consider here, providing avenues for future investigation.

Let us now turn to a detailed discussion of the evolution outlined above.
We first review the phase transition dynamics in Sec.~\ref{sec:HNEDE}  following~\cite{Garny:2024ums}, and turn to the novel aspects from incorporating the dark matter multiplet $\chi$ in  Sec.~\ref{sec:dm_dr_interaction}, proceeding to the impact of the mass splitting in Sec.~\ref{sec:mass_splitting} and the DRMD effect in Sec.~\ref{sec:decoupling}.

\subsection{Hot new early dark energy}\label{sec:HNEDE}

Before introducing a fermion dark matter multiplet, we first review the base Hot NEDE model introduced in \cite{Garny:2024ums} (building on the earlier works \cite{Niedermann:2021vgd,Niedermann:2021ijp}). Concretely, we consider the spontaneous breaking of a dark $SU(N)$ gauge group down to $SU(N-1)$, induced by non-zero dark temperature corrections to the NEDE Higgs potential 
\be\label{eq:vac_pot}
V_{\rm cl}(|\Psi|^2) = -\mu^2 |\Psi|^2 + \lambda |\Psi|^4 + V_0\,,
\ee
where $\mu$ is a mass scale, $\lambda$ is the dimensionless quartic coupling, and $V_0$ an additive constant.  

At non-zero dark temperature $T_d$, this potential will receive two types of corrections, one-loop zero-temperature quantum corrections and finite temperature thermal corrections.  Accounting for both of them, the effective potential can be written as
\begin{align}\label{Veff}
V({\psi}; T_d) = & V_0-\frac{\mu_{\rm eff}^2}{2}{\psi}^2\left(1-\frac{\psi^2}{2v^2}\right)   + B\psi^4\left(\ln\frac{\psi^2}{v^2}-\frac{1}{2}\right)  + \Delta V_\mathrm{thermal}({\psi}; T_d) \;,
\end{align}
where $\psi= \sqrt{2}|\Psi|$, $\mu_{\rm eff}$ is the effective one-loop mass providing an explicit breaking of the classical conformal symmetry, and $v$ is the scale generated by dimensional transmutation at zero temperature by the Coleman-Weinberg mechanism~\cite{Coleman:1973jx} in the limit $\mu_{\rm eff}\to 0$. The $B$ parameter is given by $B= c_1 n_A(g/2)^4/(64 \pi^2)$, where $g$ is the gauge coupling, $n_A=3(2N-1)$ is the number of gauge boson degrees of freedom  living in the coset $SU(N)/SU(N-1)$ and acquiring a mass in the phase transition, and $c_1$ is a constant  of order unity, which depends on $N$ (e.g.\ $c_1=52/45$ for $N=3$). Finally, the last term denotes the thermal corrections, which we discuss below.

The phenomenologically interesting regime in the context of the Hubble tension is when the false vacuum energy becomes cosmologically relevant after a period of strong supercooling, leading to a significant energy injection when the latent heat is released. This requires that the gauge boson mass $m_A = g v/2$ is heavy compared to the Higgs mass~\cite{Witten:1980ez}, as occurs naturally in the Coleman-Weinberg regime~\cite{Coleman:1973jx},
\begin{equation}\label{Higgsmass}
m_\psi^2=2\mu_{\rm eff}^2+8Bv^2\ll m_A^2\sim g^2v^2\,,
\end{equation}
and $m_A \gg T_d^*$, where $T_d^*$ is the dark sector temperature right before the phase transition. For typical values of those parameters, see Table~I of \cite{Garny:2024ums}. The soft breaking term $\mu_{\rm eff}$ allows us to control the amount of supercooling prior to the phase transition. In particular, as we will show below, in order to have some amount of supercooling, the effective mass must remain small $\mu^2_{\rm eff} \lesssim v^2 g^4 \sim m_\psi^2$, which is a technically natural choice due to the enhanced symmetry in the limit $\mu_\mathrm{eff}\to 0$. 

At high temperatures, the $SU(N)$ symmetry is restored by thermal corrections, while at low temperatures, when thermal corrections can be ignored, the NEDE  scalar $\Psi$ acquires a vacuum expectation value, breaking $SU(N)$ down to $SU(N-1)$. After symmetry breaking, i.e., for temperatures far below a critical temperature $T_c$ since we consider supercooling, it can be written in the form 
        \be \label{eq:vev}
            \Psi = e^{2i\pi^a\tau^a/v}
        \begin{pmatrix}
	   0 \\ \dots \\
	   \frac{v}{\sqrt{2}} + \frac{ h_\psi}{\sqrt{2}}
        \end{pmatrix} \,.
          \ee
Here we have denoted the physical mode of the dark Higgs field after the phase transition by $h_\psi$, and we have introduced the Goldstone fields related to the $2N-1$ broken generators by $\pi^a$. These Goldstone modes are absorbed by the gauge bosons acquiring mass in the unitary gauge. 

This process can be understood in greater detail by considering the $\psi$-dependence of the thermal corrections $\Delta V_\mathrm{thermal}({\psi}; T_d)$. For $\psi \ll T_d/g$, the leading thermal correction to the NEDE scalar potential is 
\begin{equation}\label{eq:barrier}
  \Delta V_\mathrm{thermal}({\psi}; T_d) \to \frac{c_0n_A}{24} g^2\,T_d^2\,\psi^2\,,
\end{equation}
where $c_0$ is a constant of order unity which depends on $N$ ($c_0=4/15$ for $N=3$) \cite{Garny:2024ums}. In particular, at sufficiently high temperatures, this term creates a thermal barrier that keeps the field in the symmetric phase at $\psi=0$. On the other hand, for large field values $\psi \gg T_d/g$, the thermal corrections to the potential are exponentially suppressed (assuming the condition in \eqref{Higgsmass} holds). This suppression becomes relevant as the dark sector temperature drops with the expansion of the Universe. Concretely, a second minimum at $\psi = v(T_d)$ appears and becomes degenerate with the symmetric minimum at the critical temperature $T_c\sim gv$ (for $\mu^2_{\rm eff} \lesssim v^2 g^4$). 
As the temperature drops further, the system enters the supercooling regime where the thermal correction in \eqref{eq:barrier} keeps the field trapped at $\psi=0$ despite the presence of a lower minimum.  By comparing the tachyonic mass term in \eqref{Veff}, $- \mu_\mathrm{eff}\, \psi^2/2$,  with the  positive thermal mass in \eqref{eq:barrier}, we find that the barrier eventually vanishes at
\begin{equation}\label{Tb}
  T_b^2 = \frac{12\mu_{\rm eff}^2}{c_0n_Ag^2}= \frac{\gamma}{\pi}g^2v^2 \,,
\end{equation}
where the supercooling parameter $\gamma$ is defined as~\cite{Niedermann:2021vgd} 
\begin{align}\label{def:gamma}
  \gamma \equiv \frac{12 \pi}{ c_0 n_A} \frac{\mu_{\rm eff}^2}{ v^2g^4}\,.
\end{align}
If $\gamma$ is less than one\footnote{We still assume that $\gamma$ is not exponentially small and in fact larger than $\sim 1/(g^2) \exp(-1/g^3)$. If this condition fails, the effective mass can be neglected and tunneling occurs as it would in the pure Coleman-Weinberg case, questioning whether the percolation condition could be satisfied.}, the phase transition occurs by bubbles of the true vacuum nucleating\footnote{We assume the bubble size to be small compared to the horizon scales at nucleation, such that the CMB is not affected by curvature fluctuations created during the phase transition~\cite{Niedermann:2020dwg,Elor:2023xbz}. As argued in \cite{Garny:2024ums,Niedermann:2021vgd}, this can be achieved with very mild restrictions on the gauge coupling or a sufficiently early phase transition that occurs before modes observable in the CMB enter the horizon. } in the supercooled regime shortly before the barrier vanishes
\begin{equation}\label{Tds} 
  T_d^* \simeq T_b = \sqrt{\frac{\gamma}{\pi}} gv \sim \sqrt{\gamma}\, T_c \,.
\end{equation}
Quantities with a star are evaluated at bubble nucleation time. In particular, for $\gamma \ll 1$, the phase transition occurs at a temperature much below the critical temperature, $T_d^* \ll T_c$, meaning in the strongly supercooled regime where the potential energy difference between the two minima is significant at the time of the phase transition. 
The released  latent heat is given more precisely by
\begin{eqnarray}\label{latent_heat}
\Delta V_*&\simeq& \left[V(\psi=v) -V(\psi=0)\right]|_{T_d=0}= T_b^4\,\frac{(3c_1+128\,\pi c_0\gamma)\,n_A}{6144\,\gamma^2}\,.
\end{eqnarray}
In the following we assume a sufficiently strong gauge coupling~\cite{Garny:2024ums}
\begin{align}\label{gauge-coupling-bound}
g \gtrsim 0.02 \,,
\end{align}
such that the phase transition leads to an almost instantaneous reheating of the dark sector, which proceeds through the loop-induced decay of the NEDE Higgs condensate into massless gauge bosons, $\psi \to A +A $, and the subsequent thermalization via gauge boson self-interactions $A+A \leftrightarrow A+A$ and (inverse) decays $ A +  A \leftrightarrow  \psi $ (see \cite{Garny:2024ums}).

Since the latent heat is released into radiation, a mix of relativistic gauge and the NEDE Higgs bosons, this causes a step-like increase in the effective number of relativistic degrees of freedom $N_\mathrm{eff}$. The dark sector contribution to $N_\mathrm{eff}$ can be quantified in terms of the temperature ratio $\xi = T_d / T_\mathrm{vis}$ as 
\begin{align}\label{def:Neff}
\Delta N_\mathrm{eff} = \frac{4}{7} \left( \frac{11}{4} \right)^{4/3} g_\mathrm{rel,d} \,\xi^4\,,
\end{align}
where $g_\mathrm{rel,d}$ and $g_\mathrm{rel,vis}$ are the numbers of relativistic bosonic degrees of freedom in the dark and visible sectors, respectively. Note that $\Delta N_\mathrm{eff}$ can be fixed in terms of $g$, $v$ and $z_*$ according to \eqref{eq:NeffNEDE}. Using that the dark sector temperature at the time of the phase transition is given by (\ref{Tds}), we can compute the small contribution to $N_{\textrm{eff}}$ \textit{before} the phase transition as
\begin{equation}\label{N_BBN}
\Delta N_\mathrm{BBN} \simeq 0.055 \, \frac{\frac{N^2+N -1}{2N-1} }{11/5}\left(\frac{\gamma}{0.01}\right)^2\frac{52/45}{c_1} \frac{ \Delta V_* / \rho_\mathrm{tot}(T_d^*)}{0.08}\,.
\end{equation}
Immediately \textit{after} the phase transition, during the fast coalescence and reheating stage, the latent heat is converted into radiation. This leads to a jump in $N_{\textrm{eff}}$
\begin{align}\label{eq:step1}
\Delta N_{\rm BBN} \to  \Delta N_{\rm NEDE} =  \frac{5 \left(3c_1 + 128\, c_0\,\pi \,\gamma\right) n_A}{1024\, \pi^2 g_\mathrm{rel,d}^* \gamma^2} \, \Delta N_{\rm BBN} \,,
\end{align}
where $g_\mathrm{rel,d}^*$ is the effective number of relativistic dark sector species just before the transition time (see Table~II of \cite{Garny:2024ums}).
We see that a strongly supercooled Hot NEDE phase transition with $\gamma \ll1$ leads to a strong reheating of the dark sector after BBN and a sharp step-like increase in $\Delta N_\mathrm{eff}$. Combining~\eqref{N_BBN} and~\eqref{eq:step1}, the parameter $\gamma$ drops out and we recover~\eqref{eq:NeffNEDE} (up to ${\cal O}(1)$ numerical prefactors).

Due to the gauge interactions, the dark radiation is tightly coupled with itself, which leads to a fluid-like behavior with vanishing higher moments, providing a natural realization of the SIDR scenario. Once the dark temperature drops below the mass $m_\psi$ of the NEDE boson, it becomes non-relativistic and decays into $2[(N-1)^2-1]$ degrees of freedom associated with the massless gauge bosons. This corresponds to a second smaller reheating of the dark sector, causing another increase in $N_{\textrm{eff}}$ (while conserving entropy, in contrast to the phase transition). This second step occurs at a redshift $z_t$ implicitly defined through $T_d(z_t) = m_\psi \sim g^2 v$. We thus have
\begin{align}\label{eq:z_t}
\frac{1+z_*}{1+z_t}= \frac{T_d^\mathrm{*,after}}{m_\psi} \simeq \frac{c_\circ}{g}\,.
\end{align}
where  $T_d^{*,\mathrm{after}}\sim{\cal O}(gv)$ is the temperature after the dark sector reheating at $z=z_*$. The last equality assumed $\gamma \ll 1$ and introduced an $N$-dependent factor that evaluates to $c_\circ=(5/91)^{1/4}\sqrt{6}\simeq 1.19$ for $N=3$~\cite{Garny:2024ums}.  
Using that the total entropy of the fluid is conserved, we
derive the step size as
\begin{align}\label{eq:N_IR}
\Delta N_{\rm NEDE} \to  \Delta N_{\rm IR} = \Delta N_{\rm NEDE} \, (1+r_g)^{1/3} \,,
\end{align}
where $r_g=(2N (N-2))^{-1}$ is the relative change in the number of relativistic degrees of freedom. Although being a prediction of our model, this second step led to only a minor improvement of the fits to the observables \cite{Garny:2024ums}.

SIDR models have been considered as proposed solutions to the Hubble tension for a while. However, it is well known that they can only provide a small alleviation of the Hubble tension, and even then, they require a too-high $\Delta N_{\textrm{eff}}\simeq 0.5$, which is excluded by BBN. This is where the Hot NEDE model steps in. The first step in $N_{\textrm{eff}}$ \textit{after} BBN in \eqref{eq:step1} provides the large initial value of $\Delta N_{\textrm{eff}}\simeq 0.5$ without being in conflict with BBN constraints. In \cite{Aloni:2021eaq}, an SIDR model with a step in $N_\mathrm{eff}$ similar to the second step in \eqref{eq:N_IR} was proposed. However, the model of \cite{Aloni:2021eaq} does not feature a first step and is therefore ruled out by BBN unless UV completed with a first step as in Hot NEDE\footnote{An alternative mechanism for generating the large initial $N_{\textrm{eff}}$ through thermalization with neutrinos was considered in \cite{Aloni:2023tff,RoyChoudhury:2020dmd}, and further mechanisms have been discussed, e.g.\ via domain wall~\cite{Ferreira:2022zzo} or particle decays~\cite{Ichikawa:2007jv,Fischler:2010xz}, see also~\cite{Sobotka:2023bzr} for a recent analysis.}. Also note that the model considered in this work is qualitatively different from scenarios for which the $\Delta N_{\rm eff}$ step generates a relative feature between the large and small-scale multipoles of the CMB \cite{Aloni:2021eaq,Schoneberg:2022grr,Schoneberg:2023rnx}. In our case, the first $\Delta N_{\rm eff}$ step \eqref{eq:step1} acts to reheat the dark radiation sector well before recombination and after BBN (recalling that the second small step \eqref{eq:N_IR} has only very little impact).

In this way, Hot NEDE, featuring the two-step sequence, $\Delta N_\mathrm{BBN} \to \Delta N_{\rm NEDE} \to  \Delta N_{\rm IR}$, can be viewed as a UV completion of the (stepped) SIDR model. As we will see below, when we add a fermion dark matter multiplet to the model, Hot NEDE also features a decoupling between dark radiation and dark matter, which leads to a full resolution of the Hubble tension. This emphasizes the importance and power of detailed realistic model building.

\subsection{Interacting dark matter and dark radiation}
\label{sec:dm_dr_interaction}

\begin{figure}[!t]
  \centering

  \begin{minipage}[t]{0.25\textwidth}
    \centering
    \includegraphics[width=\linewidth]{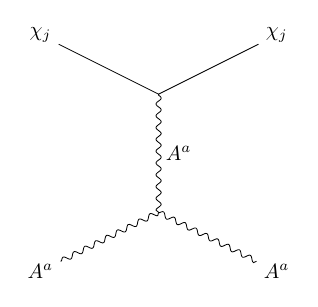}
    \par\vspace{0.5ex}
    (a) t-channel Compton scattering
  \end{minipage}
  \hspace{0.01\textwidth}
  \begin{minipage}[t]{0.25\textwidth}
    \centering
    \includegraphics[width=\linewidth]{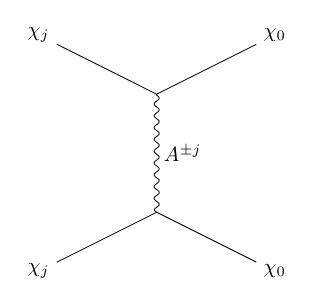}
    \par\vspace{0.5ex}
    (b) t-channel dark matter conversion
  \end{minipage}
  \hspace{0.01\textwidth}
  \begin{minipage}[t]{0.25\textwidth}
    \centering
    \includegraphics[width=\linewidth]{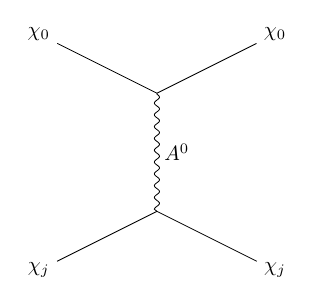}
    \par\vspace{0.5ex}
    (c) t-channel dark matter scattering
  \end{minipage}
  \caption{Dominant processes relevant for dark matter–dark radiation decoupling. 
(a) t-channel Compton scattering within the $SU(N-1)$ gauge sector maintains tight coupling between dark matter and dark radiation as long as the associated drag rate satisfies $\Gamma_\mathrm{drag} \gg H$; 
(b) t-channel conversion is relevant for chemical equilibrium between light and heavy dark matter states if the corresponding conversion rate satisfies $\Gamma_\mathrm{conv} \gg H$; 
(c) t-channel scattering between charged and neutral dark matter states establishes kinetic equilibrium within the $\chi$ sector.}
  \label{fig:Feynman}
\end{figure}

In this work, we consider an extension of the Hot NEDE model, adding also a fermion $\chi$ charged under $SU(N)$, similarly to the visible sector. Gauge and Lorentz symmetries automatically ensure stability (similarly as for the electron), making $\chi$ a candidate to compose part of dark matter. For concreteness, we assume that a fraction\footnote{In the phenomenological sections, we will use the conventional notation $f_\mathrm{idm}=f_\chi$. } $f_\chi$ of dark matter is a Dirac fermion $\chi$ with mass $m_\chi$ that transforms in the fundamental representation. Its Lagrangian reads
\be
  {\cal L} = \bar\chi (i\gamma^\mu D_\mu-m_\chi)\chi\,,
\ee
where $\chi = (\chi_1, \dots, \chi_N)$ labels the $N$ Dirac spinor components of the gauge multiplet, and with $\gamma^\mu$  being the Dirac matrices.
We further assume that $m_\chi$ is far above the dark sector temperature before and after the phase transition, such that it is sufficiently non-relativistic in the epoch between BBN and recombination, with conserved total number density when adding up the abundances of all $N$ components.

The gauge interaction of dark matter has several consequences.
Most importantly, it implies that  dark matter interacts with the dark radiation via Compton scattering. This leads to a momentum exchange, and a corresponding change in the dark matter perturbations.
More precisely, the interaction rate of dark matter with dark radiation is determined by the dark Compton scattering process of $\chi$ with massless gauge bosons, $\chi A\to \chi A$. The non-Abelian gauge bosons' self-interaction enhances the scattering cross section significantly by allowing for a $t$-channel contribution, that is absent in the Abelian
case~\cite{Moore:2004tg,Buen-Abad:2015ova,Buen-Abad:2017gxg,Rubira:2022xhb}. Similarly to baryon-photon interactions, this interaction leads to a drag force between the interacting dark matter and dark radiation
components~\cite{Buen-Abad:2015ova,Cyr-Racine:2015ihg,Chacko:2016kgg}. If the interaction rate is large enough (compared to the Hubble rate), both approach the tightly coupled limit~\cite{Buen-Abad:2017gxg,Chacko:2016kgg}.

While the gauge interactions of the dark matter multiplet are discussed in more detail in Appendix~\ref{sec:thermal}, we summarize the key results here for convenience. Before the phase transition, in the unbroken $SU(N)$ phase, we have $N^2-1$ massless gauge bosons $A^A_\mu$ transforming in the adjoint representation of $SU(N)$ with $A =1,\dots ,N^2-1$. After the phase transition, when $SU(N)$ has been spontaneously broken to $SU(N-1)$, we have the $(N-1)^2-1$ massless gauge bosons $A^a_\mu$ transforming in the adjoint of $SU(N-1)$ with $a=1,\dots,(N-1)^2-1$, and in addition $N-1$ massive (and non-relativistic) gauge boson pairs $A^{\pm j}_\mu$ with $j=1,\dots,N-1$ and mass $m_{A^\pm} =gv/2$ transforming in the fundamental representation of $SU(N-1)$, as well as a massive gauge boson $A^0_\mu$ with mass $m_{A^0_\mu}=gv\sqrt{(N-1)/(2N)}$ being a singlet under $SU(N-1)$. Moreover, we distinguish between the $N-1$ components $\chi_j=(\chi_1,\dots,\chi_{N-1})$ that transform under the remaining unbroken $SU(N-1)$ and the neutral component $\chi_0\equiv \chi_N$.

After symmetry breaking, only the massless $SU(N-1)$ gauge bosons $A_\mu^a$ remain in the dark radiation bath (alongside the relativistic Higgs $h_\psi$).
The singlet state $\chi_0$ of the dark matter multiplet does not feature direct renormalizable couplings to them. While a coupling of $\chi_0$ to the $A^0$ exists, the latter does not couple to a pair of massless gauge bosons, precluding also a $t$-channel scattering of $\chi_0$ with massless gauge bosons via $A^0$ exchange. Thus, at leading order in the gauge coupling, only the $\chi_j$ states of the dark matter multiplet have an interaction $\chi_j A^a\to \chi_j A^a$ with the massless gauge bosons.
It results from the $SU(N-1)$ gauge interactions that dominantly proceed via the $t$-channel diagram in Fig.~\ref{fig:Feynman}(a), plus an additional $t$-channel diagram with an $A^0$. The latter, due to its mass, is however
relatively suppressed at low temperatures $T\ll m_{A^0}$ after the phase transition. The reason is that the $t$-channel propagator is
enhanced for massless gauge bosons, for which it is regulated by the thermal mass of order $gT$, which is much smaller than the
gauge boson mass $m_{A^0}\sim gv$. Altogether, this means the interaction rate of the charged dark matter states, $\chi_j$, with the dark radiation bath coincides with that computed for an unbroken $SU(N-1)$ gauge theory. The rate entering the drag force is thus given by~\cite{Moore:2004tg,Buen-Abad:2015ova,Buen-Abad:2017gxg,Rubira:2022xhb}
\be \label{drag}
  \Gamma_{\chi_jA_\mu^a} = \frac{\pi\alpha_d^2}{18m_\chi}\eta_d T_d^2\big[\ln(\alpha_d^{-1})+d_0\big]\,,
\ee
where $\alpha_d = g^2/(4\pi)$ is the dark fine structure constant, $\eta_d=2((N-1)^2-1)$ is the number of degrees of freedom of the massless gauge bosons in the dark radiation bath,
and $d_0\simeq -1.34-\ln(N-1)$ for $SU(N-1)$~\cite{Rubira:2022xhb}\footnote{Here the rate is computed with respect to physical instead of conformal time. The latter is obtained by
multiplying the result with $a$. Furthermore we use a different sign convention for the drag rate as compared to~\cite{Rubira:2022xhb}.}. 

A characteristic feature of non-Abelian dark sectors is that, due to the $t$-channel enhancement, the ratio $\Gamma_{\chi_jA_\mu^a}/H$ is constant during radiation domination, and only slowly decreases afterwards, such that the drag force and the resulting tight coupling between IDM and SIDR can be active over a wide range of redshifts. Phenomenologically this can lead to a suppression of the matter power spectrum over a wide range of scales, allowing for tests with various large-scale structure data sets~\cite{Buen-Abad:2015ova,Buen-Abad:2017gxg,Rubira:2022xhb,Hooper:2022byl,Mazoun:2023kid,Euclid:2024pwi,SPT:2024roj}. However, within the model considered here, the IDM-SIDR interaction is effectively switched off at a certain point. This feature of the model is intimately related to the phase transition and the underlying gauge symmetry (breaking), and we turn to this point now.

\subsection{One-loop dark matter mass splitting}
\label{sec:mass_splitting}

\begin{figure}[!t]
  \centering
  \includegraphics[width=0.4\textwidth]{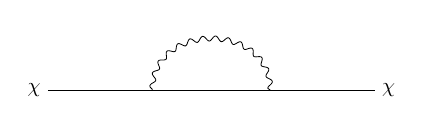}
  \caption{Self-energy corrections to the dark matter fermion propagator induce the mass splitting between neutral and charged dark matter states after a subset of the gauge bosons in the loop acquire mass via spontaneous symmetry breaking.}
  \label{fig:self-energy}
\end{figure}

The dark matter is affected by the symmetry breaking due to its gauge interaction.
Prior to the phase transition, all $N$ components of the dark matter multiplet are degenerate in mass due to the unbroken $SU(N)$ symmetry. After the transition, one-loop corrections to the self-energy of $\chi$ lead to a mass splitting (see Fig.~\ref{fig:self-energy}). Defining as usual the physical mass as the pole of the full propagator, the one-loop correction to the physical mass $m$ can be obtained from the tree-level mass $m_0$ by solving (with slight abuse of notation, see Appendix \ref{sec:splitting} for the precise meaning)
\be
\left[\slashed{p} -m_0 -\Sigma(\slashed{p})\right]|_{\slashed{p}=m}=0\,,
\ee
and computing the self-energy $\Sigma$ to one-loop order. The $N-1$ components $\chi_j\equiv (\chi_1,\dots,\chi_{N-1})$ that transform under the remaining unbroken $SU(N-1)$ remain degenerate in mass, but the component $\chi_0\equiv \chi_N$ that is neutral under $SU(N-1)$ is slightly lighter, with mass splitting 
\be \label{eq:Deltam}
  \Delta m = c_\Delta\frac{g^3v}{32\pi}\,.
\ee
We illustrate that split in blue in Fig.~\ref{fig:spectrum}.
Here $c_\Delta$ is a numerical factor of order unity (see Appendix \ref{sec:splitting}). For the case $N=3$ it is $c_\Delta=1+1/\sqrt{3}$. As we see below, this loop-induced mass splitting leads to major phenomenological consequences.

\subsection{Dark radiation matter decoupling}
\label{sec:decoupling}

The radiative mass splitting $\Delta m$ implies that, effectively, dark matter stops interacting with the dark radiation for dark sector temperatures $T_d \ll \Delta m$, as we  discuss below. 
We find this feature to be important for obtaining good agreement with CMB and LSS data for values of $H_0$ consistent with local measurements, and emphasize that this central prediction is  a natural consequence of dark matter being gauged under $SU(N)$. As mentioned before, we refer to this process as dark radiation matter decoupling (DRMD). 

Let us first discuss the main features of DRMD, before turning to a more complete discussion of the underlying assumptions. As long as the temperature $T_d$ of the dark sector is (far) above $\Delta m$, all $N$ states of the $\chi$ multiplet are equally populated. However, once $T_d\ll \Delta m$, the thermal abundance of the heavier charged components $\chi_j$ becomes Boltzmann-suppressed relative to the lighter one. Since the lighter component $\chi_0$ is neutral under the remaining $SU(N-1)$, it does (practically) not interact with the dark radiation. 

The overall $\chi$ abundance, parametrized by its constant fraction $f_\chi$ of the total dark matter density, is distributed among charged and neutral states, 
\be\label{eq:fchisplit}
  f_\chi=f_\mathrm{\chi_j}(z)+f_\mathrm{\chi_0}(z)\,,
\ee
in a redshift-dependent way. Here $f_\mathrm{\chi_j}(z)$ stands for the sum over all $N-1$ charged components of $\chi$, and $f_\mathrm{\chi_0}(z)$ for the neutral component. Directly after the phase transition $T_d\gg\Delta m$, and the charged and neutral states are occupied according to equipartition, $f_\mathrm{\chi_0}(z)/f_\chi\simeq 1/N$
and $f_\mathrm{\chi_j}(z)/f_\chi\simeq (N-1)/N$. Once $T_d\ll \Delta m$, the neutral state is preferred, and we expect $f_\mathrm{\chi_j}(z)/f_\chi\ll 1$. Indeed, in (Saha) equilibrium the ratio is (using $\Delta m, T_d\ll m_\chi$)
\begin{align}\label{eq:ratio}
\left(\frac{f_\mathrm{\chi_j}(z)}{f_\chi}\right)_\text{eq}  = \frac{(N-1)\mathrm{e}^{-\Delta m / T_d } }{(N-1)\mathrm{e}^{-\Delta m / T_d } + 1} \to (N-1)\,\mathrm{e}^{-\Delta m / T_d }\,,
\end{align}
where the latter limit applies for $T_d\ll \Delta m$, while equipartition is recovered in the opposite case. A suppression of the $\chi_j$ abundance implies that the effective drag rate $\Gamma_\text{drag}$ for dark radiation is reduced relative to $\Gamma_{\chi_jA_\mu^a}$ from~\eqref{drag} as
\begin{align}\label{eq:Gamma_drag}
\Gamma_\mathrm{drag} = \frac{f_{\chi_j}(z)}{f_\chi} \Gamma_{\chi_jA_\mu^a}\,.
\end{align}
While the ratio $\Gamma_{\chi_jA_\mu^a}/H$ is approximately constant, $\Gamma_\mathrm{drag}/H$ rapidly decreases once the neutral state is preferred, and if the ratio drops below unity the dark radiation cannot maintain tight coupling to the non-relativistic $\chi$ components, giving rise to DRMD.

Let us first estimate the redshift range relevant for DRMD, postponing a more detailed discussion for the moment. As a starting point we take the redshift $z_\text{stop}$ for which the dark sector temperature drops below the mass splitting, i.e.~$T_d(z_\text{stop})=\Delta m$. Furthermore, we consider a redshift $z_\text{dec}$ defined by the requirement that dark matter and dark radiation perturbations cease to be tightly coupled to each other. Provided this occurs due to the reduction of the drag rate, the redshift is set by the time when $\Gamma_\mathrm{drag}/H$ drops below unity, as we assume below for definiteness. Once more, see Fig.\,\ref{fig:evolution} for an illustration.

To provide a  quantitative estimate, we assume that the fraction of charged states is given by the equilibrium value~\eqref{eq:ratio}, and parametrize the expansion history conveniently as
\begin{align}
a^2 H = (a^2 H)|_\mathrm{after}\,\sqrt{ 1+ \frac{\Omega_m}{\Omega_\mathrm{rad} } \frac{1}{1+z} }\,,
\end{align}
where the subscript `after' stands for evaluation after the phase transition at $z_*$ when the dark sector reheating has been completed (during radiation domination). It is given by
\begin{align}
\label{eq:aH_after}
(a^2 H)|_\mathrm{after} = \frac{\pi}{3\sqrt{10}} \frac{1}{M_\mathrm{pl}} \sqrt{g_\mathrm{rel,vis}+g_\mathrm{rel,d}\xi_\mathrm{*,after}^4}\, a^2 T_\mathrm{vis}^2\,.
\end{align}
For the redshift $z_\mathrm{stop}$ at which the exponential suppression kicks in, we read off from~\eqref{eq:ratio}, 
\begin{align}\label{eq:z_stop}
\frac{1+z_\mathrm{stop}}{1+z_*} =   (1+r_g)^{-1/3}   \frac{\Delta m }{T_d^{*,\mathrm{after}}} \,,
\end{align}
where the factor $(1+r_g)^{-1/3}$ takes into account the  mass-threshold effect that occurs at $z_t$. Noticing from \eqref{eq:mass_hierarchy} that $m_\psi/\Delta m \sim  (32\pi)/g$, the different redshifts satisfy the hierarchy $z_\mathrm{stop} \ll z_t \ll z_*$. We can make this more explicit by taking the ratio between \eqref{eq:z_t}  and \eqref{eq:z_stop}, 
\begin{align}\label{def:z_stop_z_t}
\frac{1+z_t}{1+z_\mathrm{stop}} =(1+r_g)^{1/3}  \frac{m_\psi}{\Delta m} \simeq \frac{c_\Box}{g}\,,
\end{align}
where $c_\Box$ is another $N$-dependent factor that for $N=3$ evaluates to $c_\Box  = 4 \times 6^{1/6} \, 7^{1/3} \, \sqrt{13}\, (3+\sqrt{3}) \simeq 7.86  $ (assuming $\gamma \ll1$ and using \eqref{Higgsmass} and \eqref{eq:Deltam}). In other words, we obtain from \eqref{eq:z_t} and \eqref{def:z_stop_z_t} the redshift hierarchy 
\begin{align}\label{eq:z-ladder}
z_* \simeq \frac{c_\circ}{g} \, z_t \simeq \frac{c_\Box \,c_\circ}{g^2} z_\mathrm{stop} \,,
\end{align}
which is controlled by the gauge coupling $g$. The condition $\Gamma_\mathrm{drag} = H$ is satisfied even later for 
$z_\mathrm{dec} < z_\mathrm{stop}$.
Neglecting the second term under the square root in \eqref{eq:aH_after}, taking $g_\mathrm{rel,vis}=3.38$, and using \eqref{def:Neff} together with \eqref{eq:N_IR}, we obtain from \eqref{eq:Gamma_drag} (a more explicit derivation is performed in Sec.~\ref{sec:perts})

\begin{align}\label{eq:z_ratio}
\frac{1+z_\mathrm{stop}}{1+z_\mathrm{dec}}
\simeq 28 + \ln\left[ \frac{16 \pi^2\alpha^2_d}{(0.1)^4}\,\big(\ln(\alpha_d^{-1})+d_0\big)\,\left( 1+ \frac{\Omega_m}{\Omega_\mathrm{rad} } \frac{1}{1+z_\mathrm{dec}} \right)^{-1/2}\sqrt{\Delta N_\mathrm{IR}}\, \left(\frac{\mathrm{GeV}}{m_\chi}\right) \right]\,,
\end{align}
where we have substituted $N=3$ for illustration. Anticipating our numerical results, we are interested in cases where the decoupling happens close to matter-radiation equality, i.e.\ $z_\mathrm{dec} \sim z_\mathrm{eq}$. Therefore, we can estimate $\Omega_m/\Omega_\mathrm{rad} \sim z_\mathrm{dec}$, and thus ignore the $z_\mathrm{dec}$ inside the log term. Moreover, we find that $\Gamma_\mathrm{drag}\gg H$ is initially satisfied for realistic values of the dark matter mass. To be explicit, we obtain a very weak upper bound on the dark matter mass:
\begin{align}
m_\chi \lesssim (g/0.1)^4\, 10^{12} \,\mathrm{GeV}  \,.
\end{align}
Later, we will see that the DRMD model is efficient at addressing the Hubble tension if $z_\mathrm{dec} \sim 1000 $. Using \eqref{eq:z_ratio} and \eqref{eq:z-ladder} together with $g \lesssim 0.1$, this implies the lower bound $z_* \gtrsim 10^6$. In other words, the decoupling mechanism constrains the phase transition to occur in the redshift window 
\begin{align}\label{z_range}
10^6 \lesssim z_* \lesssim   10^9\,,
\end{align}
which slightly tightens the range from~\cite{Garny:2024ums}.

\medskip

Let us now discuss some refinements and underlying assumptions.
The baseline picture followed here assumes kinetic and chemical equilibrium in the dark sector during DRMD at the background level, as well as a tight coupling between the perturbations of charged and neutral states. Kinetic equilibrium within the dark sector at a common temperature $T_d$ for $\chi_j$, $\chi_0$ and $A_\mu^a$ relies on the exchange of kinetic energy via elastic scatterings, that are efficiently mediated by gauge interactions. An example is the $2\to 2$ process $\bar\chi_j\chi_0\to \bar\chi_0\chi_j$, which can proceed through the $s$ and $t$-channel exchange of $A^{\pm j}$ and $A^0$, respectively [see Fig.~\ref{fig:Feynman}(c)]. At the same time, apart from momentum, the heavier dark matter states $\chi_j$ also exchange energy with the massless gauge bosons in the radiation bath via the scattering $\chi_j A^a \to \chi_j A^a$, mediated by a $A^a$ [see Fig.~\ref{fig:Feynman}(a)]. On the other hand, chemical equilibrium in the conversion among charged and neutral states can be established via the inelastic process $\bar\chi_j\chi_j\to \bar\chi_0\chi_0$. It can proceed via the $A^0$ gauge boson in the $s$-channel and via $t$-channel diagrams with exchange of $A^{\pm j}$ [see Fig.~\ref{fig:Feynman}(b)]. The latter are enhanced due to the $t$-channel propagator being of the order of the momentum transfer (see Appendix~\ref{sec:conversion}), while the $s$-channel propagator is suppressed by the dark matter mass. Finally, at the level of perturbations, tight coupling between $\chi_j$ and $\chi_0$ relies on momentum exchange, with relevant processes being the same elastic scattering $\bar\chi_j\chi_0\to \bar\chi_0\chi_j$ as for kinetic equilibration at the background level. Since the relevant rates involving in particular processes with two non-relativistic particles such as $\bar\chi_j\chi_0\to \bar\chi_0\chi_j$ and $\bar\chi_j\chi_j\to \bar\chi_0\chi_0$ are potentially subject to large (Sommerfeld-)corrections, we defer a detailed analysis of the various processes to future work. For example, it would be interesting to investigate whether DRMD could occur due to the breakdown of tight coupling between $\chi_0$ and $\chi_j$ instead of dark radiation and $\chi_j$, as assumed above. Further points are residual interactions of $\chi_0$ with $SU(N-1)$ gauge bosons that could be induced by loop effects (such as electric or magnetic dipole interactions), as well as a breakdown of Saha equilibrium among the $\chi_0$ and $\chi_j$ states, similarly to what occurs for the later stages of recombination in the visible sector. Finally, it would be worthwhile to consider model extensions, e.g.\ by adding massless fermion species $q$ charged under $SU(N)$ to the model, that  yield a (small) extra contribution to the SIDR component, and  mediate further conversion processes between $\chi_j$ and $\chi_0$ states, allowing also for $1\to 3$ decays such as $\chi_j\to\chi_0 q^j \bar q^0$ that are effective in depleting the charged states down to arbitrarily low temperatures.

Altogether, while a more detailed investigation is of interest, we expect the basic picture of DRMD to be robust: for $T_d\ll\Delta m$ the charged states $\chi_j$ become depleted, such that dark radiation stops interacting efficiently with the (then dominantly neutral) $\chi$ component at some point. Thus, tight coupling between SIDR and dark matter terminates, increasing the sound velocity for dark radiation while non-Abelian self-interactions keep it from starting to free-stream. As we see below these are important features at the perturbation level for making the model consistent with CMB and BAO data while allowing for a reduction of the sound horizon and thus a large value of $H_0$, being generically provided by the model discussed here. Specifically, the redshift $z_\text{stop}$ at which DRDM dynamics starts is independent of the uncertainties discussed above, while the redshift $z_\text{dec}$ and the estimate provided in~\eqref{eq:z_ratio} could differ slightly depending on which of the equilibrium conditions discussed above breaks down first.

\medskip

Overall, guided by fundamental principles, we have discussed a microscopic model governed by a dark non-Abelian gauge symmetry $SU(N)$, spontaneously broken to $SU(N-1)$ by a Higgs mechanism. In the Coleman-Weinberg regime, supplemented by a small technically natural explicit breaking that controls the exit from supercooling, this setup gives rise to a supercooled phase transition that can create an SIDR component with sizeable $\Delta N_\text{eff}$ after BBN. Independently, the same symmetry breaking induces a radiative mass splitting $\Delta m$ of a heavy fermion charged under the gauge symmetry. This provides a tight coupling of SIDR to an interacting dark matter component that terminates once the dark sector temperature drops below $\Delta m$. Both of these features, (i) the reheating of the dark sector after BBN and (ii) a subsequent loop-induced decoupling, play a key role in the way the model addresses the Hubble tension. We now discuss a setup allowing us to efficiently track the evolution of perturbation modes relevant for a comparison to CMB data.

\section{Evolution of perturbations}
\label{sec:pheno}

In this section we work out the evolution of perturbations within the dark sector  for the microscopic Hot NEDE model discussed above, and connect its fundamental parameters to those entering the perturbation equations (Sec.~\ref{sec:perts}). 
For that purpose we extend the setup from~\cite{Garny:2024ums} by the interacting dark matter component $\chi$.
Furthermore, we present a simplified description in Sec.~\ref{sec:DRMD} that we refer to as \textit{Dark Radiation Matter Decoupling Model} (DRMD). It captures the essential DRMD dynamics, being applicable if {\it (i)} the dark sector phase transition occurs sufficiently early such that perturbation modes observed in the CMB were still outside of the horizon at that time  (but still after BBN), requiring roughly $10^6\lesssim z_* \lesssim 10^9$ (in accordance with \eqref{z_range}), and {\it (ii)} the mass-threshold effect at $z_t$ due to the Higgs boson becoming non-relativistic is negligible, which is the case in the large-$N$ limit of $SU(N)$.

\subsection{Dark matter and dark radiation perturbations} \label{sec:perts}

We describe perturbations in the dark sector by adding two additional components to the $\Lambda$CDM model, one accounting for dark radiation (DR) and one for interacting dark matter ($\chi$). Dark radiation includes the massless gauge bosons, as well as the Higgs as long as it is relativistic. As discussed in~\cite{Garny:2024ums}, non-Abelian self-interactions of gauge bosons as well as interactions of the Higgs allow  for a fluid description in terms of the density contrast $\delta_\text{DR}$ and velocity divergence $\theta_\text{DR}$, suppressing anisotropic stress and higher multipoles. The interacting dark matter component includes the neutral $\chi_0$ and charged $\chi_j$ states, assumed to be tightly coupled to each other (see Sec.~\ref{sec:decoupling}) and being described by a cold (i.e.~pressureless) fluid, in line with our assumption $m_\chi\gg T_d$, with density contrast $\delta_\chi$ and velocity divergence $\theta_\chi$. As discussed above we assume that $\chi$ contributes as a constant fraction $f_\chi$ of the total dark matter density, while the remaining part is taken to be standard CDM (a natural possibility is a further particle species being a singlet under $SU(N)$).\footnote{Since we are interested in a scenario with strong coupling between $\chi$ and DR, the case $f_\chi = 1$ is very constrained by data since it generates strong dark acoustic oscillations \cite{Buen-Abad:2015ova,Chacko:2016kgg,Buen-Abad:2017gxg}.} We further use~\eqref{eq:fchisplit} for the redshift-dependent relative abundances in the $\chi_0$ and $\chi_j$ states, and moreover assume standard adiabatic initial conditions for all species on superhorizon scales.

The perturbations of the interacting dark matter component in synchronous gauge satisfy (following the notation in~\cite{Ma:1995ey,Buen-Abad:2017gxg})
\begin{subequations}
\label{eq:IDM}
\begin{align}
\delta_\chi^\prime + \left( \theta_\chi + \frac{h^\prime}{2} \right) &= 0 \,, \label{eq:delta_idm}\\
\theta_\chi^\prime + \mathcal{H} \theta_\chi &= \mathcal{G} \Delta^\prime \,,\label{eq:theta_idm}
\end{align}
\end{subequations}
where
\begin{align}
\Delta' = 
\theta_\mathrm{DR}-\theta_\chi\,,
\end{align}
defines the velocity slip, $h$ is the spatial trace of the metric fluctuation, and $\mathcal{G}$ is a time-dependent function quantifying the momentum transfer between dark matter and dark radiation, related to the drag rate~\eqref{eq:Gamma_drag}, see below for details. Furthermore, primes denote derivatives with respect to conformal time $\tau$ and $\mathcal{H} = a H$. 

Perturbations in the dark radiation fluid are governed by
\begin{subequations}
\label{eq:DR}
\begin{align}
\delta_\mathrm{DR}^\prime + \left( 1+ w_{\rm DR} \right)\left(\theta_{\rm DR} + \frac{h^\prime}{2}\right)  + 3 \mathcal{H} \left(c_s^2 - w_{\rm DR} \right) \delta_\mathrm{DR}&= 0\,, \label{eq:delta_DR}\\
\theta^\prime_\mathrm{DR} - \frac{k^2 c_s^2}{1 + w_{\rm DR}} \d_{\rm DR} + \mathcal{H} \left(1-3 c_s^2  \right) \theta_\mathrm{DR} &= -\mathcal{G}R \Delta^\prime \label{eq:theta_DR}\,,
\end{align}
\end{subequations}
where we have defined
\begin{align}\label{eq:R}
R = \frac{1}{1+w_\mathrm{DR}}
 \frac{\rho_\chi}{\rho_\mathrm{DR}}\,,
\end{align}
and 
\begin{align}
c_s^2= w_\mathrm{DR} - \frac{w_\mathrm{DR}^\prime}{3 \mathcal{H} (1+w_\mathrm{DR})}\,,
\end{align}
is the fluid sound velocity.
The factor $R$ arises because we impose conservation of the combined fluid with velocity divergence $( 1+ w_\mathrm{eff}) \rho \,\theta  \equiv (1+w_\mathrm{DR})\, \rho_\mathrm{DR} \,\theta_\mathrm{DR} + \rho_\chi \,\theta_\chi$, where $\rho=\rho_\mathrm{DR}+\rho_\chi$ and $w_\mathrm{eff}$ is the effective equation of state. Note that we allow for an equation of state $w_\mathrm{DR} \neq 1/3$, which arises because the radiation fluid after the phase transition consists of a mixture of massless gauge bosons and the Higgs that becomes non-relativistic at some point. As mentioned above, the vanishing of higher multipole moments of dark radiation perturbations is a consequence of self-interactions within this sector.  Microscopically, those are  realized through the gauge interactions of  $SU(N)$  [and the residual $SU(N-1)$] before (and after) the phase transition, such as $\psi\leftrightarrow AA$ and $AA\leftrightarrow AA$~\cite{Garny:2024ums}. 

We now describe the cosmological evolution of the dark sector fluids during all epochs resulting from the microscopic model (see Fig.\,\ref{fig:evolution}) in several steps: the pre-phase-transition stage for $z>z_*$, the phase transition at $z=z_*$, and the subsequent evolution, which contains the mass-threshold effect of the Higgs at $z=z_t$ and the later decoupling of $\chi$ and DR components at $z_\text{stop}\gtrsim z\gtrsim z_\mathrm{dec}$.

\subsubsection{\texorpdfstring{Before the phase transition ($z > z_{*}$)}{Before the phase transition (z > z*)}}

Before the phase transition, all $SU(N)$ gauge bosons are massless and as a consequence all $\chi$ dark matter degrees of freedom are interacting. Moreover, the (subdominant) dark radiation fluid, composed of $2(N^2-1)$ massless gauge bosons and $2N$ highly relativistic Higgs bosons, can be characterized by $c_s^2=w_\mathrm{DR}\simeq1/3$. 
We further assume that dark radiation and dark matter are tightly coupled. 
This is described by the limit $\mathcal{G} \to \infty$, which simplifies the fluid equations significantly \cite{Blas:2011rf,Buen-Abad:2017gxg}. It follows from \eqref{eq:theta_idm} that $\Delta^\prime \to 0$. We then obtain from \eqref{eq:delta_idm} and \eqref{eq:delta_DR}, 
\begin{align}\label{delta_idm}
\delta_\chi = 
\frac{3}{4}\delta_\mathrm{DR} + C^{(-)} \,,
\end{align}
where the integration constant $C^{(-)}$ is vanishing for adiabatic initial conditions. We note that this can be written equivalently as $\Delta = 0$ at leading order in $1/{\cal G}$. Taking the difference between \eqref{eq:theta_idm} and \eqref{eq:theta_DR}, it is straightforward to derive the momentum transfer
\begin{align}\label{GDelta}
\mathcal{G} \Delta^\prime = 3 c_{s,\mathrm{eff}}^2
 \left[\frac{ k^2}{4} \delta_\mathrm{DR} +  \mathcal{H}\theta \right]\,,
\end{align}
where the sound velocity of the combined fluid is
\begin{align}
c_{s,\mathrm{eff}}^2 = 
\frac{1}{3 (1+R)}\,,
\end{align}
due to  `dark matter loading'.

In practical terms, we use \eqref{eq:DR} together with \eqref{GDelta} to evolve the system. The dark matter perturbation $\delta_\chi$ is then obtained from \eqref{delta_idm}. As is well known from the baryon-photon system, in the tight-coupling limit the interaction enters only indirectly via enforcing the common sound velocity $c_{s,\mathrm{eff}}^2$ and we do not need to specify the interaction rate ${\cal G}$.

\subsubsection{\texorpdfstring{Phase transition ($z=z_*$)}{Phase transition (z = z*)}}

The first-order phase transition at $\tau=\tau_*$ takes place quasi-instantaneously on cosmological time-scales, and leads to a discontinuous increase in the DR energy density due to the latent heat $\Delta V$ being converted into radiation.
We therefore have for the background DR density
\begin{align}
{\rho}_\mathrm{DR}^{(+)} = {\rho}_\mathrm{DR}^{(-)} + \Delta V \,,
\end{align}
where in the supercooled regime $\Delta V \gg {\rho}_\mathrm{DR}^{(-)}$. Here the superscript $(\mp)$ indicates evaluation before and after the phase transition, i.e.\ $f^{(\mp)} =\lim_{\epsilon \to 0 } f(\tau_* \mp \epsilon) $ for any function $f(\tau)$.  This assumes an almost instantaneous energy conversion, which, as discussed in \cite{Garny:2024ums}, is justified if the Higgs decays into massless gauge bosons efficiently [see also \eqref{gauge-coupling-bound}]. 

This instantaneous conversion is described at the level of perturbations through the matching equations
\footnote{An intuitive understanding of the matching conditions can be gained through the following argument. Under a general linear diffeomorphism $\xi^\mu(\tau,\mathbf{k})$, the two fluid variables transform as $\delta_\mathrm{DR} \to \hat{\delta}_\mathrm{DR} =  \delta_\mathrm{DR} + 4 \mathcal{H} \xi^0$ and $\theta_\mathrm{DR} \to \hat{\theta}_\mathrm{DR} =  \theta_\mathrm{DR} - k^2 \xi^0$. Consequently, the gauge transformation $(\xi^0)^{(-)} = -  \delta_\mathrm{DR}^{(-)}/(4 \mathcal{H(\tau_*)})$ takes us to a frame where $\hat{\delta}_\mathrm{DR}^{(-)}= 0 $. Since the phase transition is triggered along a surface of constant $\rho_\mathrm{DR}(\tau, \mathbf{x})$, it now occurs simultaneously across space (at least on scales much larger than the typical bubble separation where the stochastic nature of the phase transition can be ignored). As a result, in this `trigger frame', the macroscopic fluid is homogeneously produced and thus \textit{both} $\hat{\delta}_\mathrm{DR}^{(+)}= 0 $ and $\hat{\theta}_\mathrm{DR}^{(+)} = 0$. Transforming back to the original frame, then recovers \eqref{eq:matching}. This argument assumes that ${\rho}_\mathrm{DR}^{(+)} \gg {\rho}_\mathrm{DR}^{(-)}$, meaning that only the spatial dependence in the trigger surface, i.e.\ $\delta _\mathrm{DR}^{(-)}$, is responsible for the post-phase transition perturbations.}

\begin{align}
\label{eq:matching}
\delta_\mathrm{DR}^{(+)} &= \delta_\mathrm{DR}^{(-)}\,, \quad
\theta_\mathrm{DR}^{(+)} =  - \frac{k^2}{4 \mathcal{H}_*}  \delta_\mathrm{DR}^{(-)}\,,
\end{align}
where we have made the simplifying assumption that $w_\mathrm{DR}^{(+)} = w_\mathrm{DR}^{(-)} = 1/3$. Perturbations $\delta_\mathrm{DR}^{(-)}$ in the subdominant dark radiation fluid before the phase transition thus seed perturbations in the dark sector fluid, $\delta_\mathrm{DR}^{(+)}$ and $\theta_\mathrm{DR}^{(+)}$, after the phase transition, imprinting characteristic features in the matter power spectrum for co-moving scales with wave number $k  > \mathcal{H}_*$~\cite{Garny:2024ums}. 
After the phase transition, we assume that tight coupling between the reheated dark radiation fluid and dark matter is quickly restored, i.e. $\theta_\chi \to \theta^{(+)}_\mathrm{DR}$, as a consequence of the efficient momentum transfer between both fluids (see Fig.~\ref{fig:Feynman}(a) for a microscopic process). 
% %
In summary, \eqref{eq:matching} provides the initial conditions for the dark radiation perturbations after the (almost instantaneous) reheating of the dark sector has been completed. 
The evolution equations~\eqref{eq:DR} along with~\eqref{GDelta} describing the tightly coupled limit remain valid also after the phase transition. We note that the matching conditions are not required for the simplified model described in Sec.~\ref{sec:DRMD}.

\subsubsection{\texorpdfstring{Mass-threshold effect of the Higgs ($z = z_t$)}{Mass-threshold effect of the Higgs (z = zt)}}

Another prediction of the Hot NEDE model, described in \cite{Garny:2024ums}, is a (minor) further increase in $\Delta N_\text{eff}$ (`second step'), in which the Higgs field becomes non-relativistic. This step vanishes in the large-$N$ limit of $SU(N)$, and was shown to have little influence on the fit to CMB data (see Fig.~8 of \cite{Garny:2024ums}) even for $N=3$. For this reason, we do not implement this second step in the simplified DRMD model (see Sec.~\ref{sec:DRMD}). We include a description of this second step here for completeness, since it is part of the cosmological history of Hot NEDE, and could hold testable predictions for future higher-precision data sets.

The post-phase-transition DR fluid consists of $2[(N-1)^2-1]$ massless gauge bosons and one massive Higgs degree of freedom in thermal equilibrium. Following~\cite{Aloni:2021eaq,Garny:2024ums}, this mixed system can be described as a single fluid  with time-dependent equation-of-state parameter
\begin{align}
w_\mathrm{DR}(x) = \frac{1}{3} - \frac{r_g}{3}\frac{\hat{\rho}(x) - \hat{p}(x)}{1+r_g\hat{\rho}(x)} \,, 
\end{align}
where $r_g^{-1}=2N (N-2)$, and $\hat{\rho}(x)$ and $\hat{p}(x)$ are expressed in terms of modified Bessel functions of the second kind,
\begin{align}
\hat{\rho}(x) & = \frac{K_1(x)}{6}x^3 + \frac{K_2(x)}{2}x^2   \,, \nonumber\\
\hat{p}(x) &= \frac{K_2(x)}{2}x^2   \,.
\end{align}
The relation between $a$ and $x = m_\psi/T_d$ can then be obtained by numerically solving 
\begin{align}
\left(\frac{xa_{t}}{a}\right)^3 = 1 + \frac{r_g}{4}\left( 3 \hat{\rho}(x)+\hat{p}(x)\right) \,,
\end{align}
which imposes entropy conservation of the combined fluid. Here $a_t \equiv 1/(1+z_t)$ corresponds to the scale factor where the Higgs becomes non-relativistic. As shown in \eqref{eq:z_t}, the redshift ratio $z_t/z_*$ is parametrically controlled by $g$. In particular, for $z_t \ll z_*$, the Higgs is highly relativistic with $x \ll 1$ and the above equations yield $w_\mathrm{DR} \simeq 1/3$ in agreement with the assumptions made for the perturbation matching in \eqref{eq:matching}.

Physically, these equations describe how the (dark) Higgs boson, when it becomes non-relativistic and annihilates, heats the dark radiation fluid, leading to an increase in $\Delta N_\mathrm{eff}$. According to \eqref{eq:N_IR}, the size of this second step is quantified in terms of $r_g$ and vanishes for $N \to \infty$. 

\subsubsection{\texorpdfstring{Dark matter decoupling $(z_{\rm stop}> z > z_{\rm dec})$}{Dark matter decoupling (z = zdec)}}

As a new feature of this  work, we include the interaction of the dark radiation fluid with dark matter.
As discussed in Sec.~\ref{sec:micro}, due to the mass splitting $\Delta m$ between the neutral and the $(N-1)$ charged components of the $\chi$ multiplet, the interaction rate receives an additional time-dependent Boltzmann suppression. As a result, the charged sector is depopulated as $T_d < \Delta m $, which in turn suppresses the interaction rate and hence $\mathcal{G}$. Since $\Delta m \sim g^3 v \ll m_\psi \ll T_d^* $, this happens after the phase transition and the second step at $z=z_t$.
To describe this stage, we cannot use the tight coupling approximations and go back to the separate set of evolution equations for DR~\eqref{eq:DR} and $\chi$~\eqref{eq:IDM}. This requires us to explicitly specify the interaction rate ${\cal G}$, which we discuss in the following.

For easier comparison with the literature, we specify the relation with the ETHOS framework~\cite{Cyr-Racine:2015ihg}. To be concrete, the parameter $\mathcal{G}$ is related to the opacity $\Gamma_{\mathrm{DR}-\mathrm{idm}}$ through $\mathcal{G} \equiv - \frac{1}{R}\Gamma_{\mathrm{DR}-\mathrm{idm}}$, which is conventionally parametrized as
\begin{align}
\Gamma_{\mathrm{DR}-\mathrm{idm}} = -\left(\Omega_\chi h^2 \right) x_\mathrm{idm}(z) \sum_{n\geq 0} a_n \left(\frac{1+z}{1+z_D} \right)^n \,.
\end{align}
Our case is captured by the choice $a_{n>0}=0$. The coefficient $a_0$ is related to the microscopic parameters in the previous section through
\begin{align}
a_0 &= \frac{a\,R}{\Omega_\chi h^2} \,\Gamma_{\chi_jA_\mu^a}\nonumber\\
&= \frac{a\,R}{\Omega_\chi h^2}  \frac{\pi\alpha_d^2}{18m_\chi}\eta_d T_d^2\big(\ln(\alpha_d^{-1})+d_0\big)\,,
\end{align}
which, except for a mass-threshold effect around $z=z_t$, is indeed constant during radiation domination because $T_d \propto 1/a$ and $R \propto a$ (as $w_\mathrm{DR}(\tau) \simeq 1/3$).
The factor $x_\mathrm{idm}(z)$ accounts for any additional time dependence.  In particular, it captures the Boltzmann suppression of the charged dark matter component $\chi_j$,
\begin{align}\label{eq:x_idm}
x_\mathrm{idm}(z)=\frac{f_\mathrm{\chi_j}(z)}{f_\chi} x_\mathrm{back}(z) = \frac{(N-1)\mathrm{e}^{-\Delta m / T_d } }{(N-1)\mathrm{e}^{-\Delta m / T_d } + 1}\, x_\mathrm{back}(z)\,,
\end{align}
where we have assumed the equilibrium ratio~\eqref{eq:ratio} and used that the number density of a single non-relativistic degree of freedom is $n_1 \simeq 2 \sqrt{2} \pi^{3/2} e^{-T_d/m} (T_d/m) ^{3/2}$. The factor $x_\mathrm{back}$ absorbs deviations of the expansion history from pure radiation domination. 
We then obtain
\begin{align}\label{def:G_0}
\mathcal{G} =  \left(\Omega_\chi h^2 \right) \frac{a_0}{R} \, x_\mathrm{idm}(z)\,.
\end{align}
Right after the phase transition, we have $\Delta m \ll T_d$ and $x_\mathrm{back} \simeq 1$, and thus
\begin{align}
\left(\frac{\mathcal{G}}{\mathcal{H}}\right)\Bigg|_\mathrm{after} \simeq \frac{N-1}{N} \left(\Omega_\chi h^2 \right) \left(\frac{a_0}{\mathcal{H} R}\right)\Big|_\mathrm{after} \,.
\end{align}
As we are interested in a situation where $(\mathcal{G}/\mathcal{H})_* \gg 1$, we can employ the tight-coupling approximation, which makes the fluid evolution initially insensitive to the precise redshift dependence of $\mathcal{G}$. Eventually, at some later redshift $z\ll  z_t$, we have  $\Delta m \gg T_d$, and \eqref{eq:x_idm} yields\footnote{A similar exponential shutoff of the DM--DR drag rate appears in SPartAcous-type models~\cite{Buen-Abad:2022kgf}.}
\begin{align}
\left(\frac{\mathcal{G}}{\mathcal{H}}\right) \simeq N\,\left(\frac{\mathcal{G}}{\mathcal{H}}\right)\Bigg|_\mathrm{after}  x_\mathrm{back}(z)\,\mathrm{e}^{-\Delta m / T_d } \,,
\end{align}
where we read off
\begin{align}
x_\mathrm{back}(z) = \frac{a_0\phantom{|_\mathrm{later}}}{a_0|_\mathrm{after}}\frac{(\mathcal{H}R)|_\mathrm{after}}{\mathcal{H}R}=(1+r_g)^{2/3}\left( 1+ \frac{\Omega_m}{\Omega_\mathrm{rad} } \frac{1}{1+z} \right)^{-1/2} \,,
\end{align}
which accounts for both the mass-threshold effect at $z=z_t$ and deviations from radiation domination.
For our later Boltzmann evolution, it is therefore useful to introduce
\begin{align}
\left[\frac{\mathcal{G}}{\mathcal{H}}\right]_\mathrm{ini}= N\,(1+r_g)^{2/3}(\mathcal{G}/\mathcal{H})\Big|_\mathrm{after}
\end{align}
as the input parameter.\footnote{We note that this assumes that the tight-coupling approximation is still valid when $\Delta m \sim T_d$. Otherwise, one needs to use the full expression in \eqref{eq:x_idm} to describe the decoupling.} It can be related to the microscopic parameters introduced in Sec.~\ref{sec:micro} through~\cite{Rubira:2022xhb}
\begin{align}
 \left[\frac{\mathcal{G}}{\mathcal{H}}\right]_\mathrm{ini} 
 &= (N-1) \frac{a_*^2  \Gamma_{\chi_jA_\mu^a}}{a_*\mathcal{H_*}} \nonumber\\
 &=  \frac{\sqrt{10}}{3} \, \frac{(N-1)((N-1)^2-1)}{\sqrt{g_\mathrm{rel,vis}+g_\mathrm{rel,d}\xi_{*,\mathrm{after}}^4}}\,\alpha_d^2
 \, \big(\ln(\alpha_d^{-1})+d_0\big) \xi_{*,\mathrm{after}}^2  (1+r_g)^{2/3}\frac{M_\mathrm{pl}}{m_\chi}
 \,,
\end{align}
where in the second line, we substituted \eqref{drag} and used (\ref{eq:aH_after}) written in conformal time.
Neglecting the second term under the square root, taking $g_\mathrm{rel,vis}=3.38$, and using \eqref{def:Neff} together with \eqref{eq:N_IR}, we obtain 
\begin{align}\label{G_ini_numerical}
\left[\frac{\mathcal{G}}{\mathcal{H}}\right]_\mathrm{ini} \simeq 1.5 \times 10^{12}\, \frac{(N-1)\sqrt{(N-1)^2-1)}}{2 \sqrt{3}}\, \frac{16 \pi^2\alpha^2_d}{(0.1)^4}\,\big(\ln(\alpha_d^{-1})+d_0\big)\sqrt{\Delta N_\mathrm{IR}}\, \left[\frac{\mathrm{GeV}}{m_\chi} \right]\,.
\end{align}
We further parametrize the full time dependence (valid for $\Delta m \gg T_d$) as
\begin{align}\label{def:G}
\mathcal{G} \simeq \mathcal{H} \left(\frac{\mathcal{G}}{\mathcal{H}}\right)_\mathrm{ini} \left( 1+ \frac{\Omega_m}{\Omega_\mathrm{rad} } \frac{1}{1+z} \right)^{-1/2} \, \exp \left( -\frac{1+z_\mathrm{stop}}{1+z} \right)\,,
\end{align}
where we use the redshift $z_\mathrm{stop}$ at which the exponential suppression kicks in defined in~(\ref{eq:z_stop}).
Requiring that ${\cal G}/{\cal H}=1$ at $z=z_\text{dec}$ yields
\begin{align}\label{eq:z_ratio2}
\frac{1+z_\mathrm{stop}}{1+z_\mathrm{dec}} & \simeq \ln\left( \left[\frac{\mathcal{G}}{\mathcal{H}}\right]_\mathrm{ini} \right) \,.
\end{align}
Substituting \eqref{G_ini_numerical} for $N=3$, we recover the numerical estimate given in~(\ref{eq:z_ratio}).

A priori, the microscopic model parameters are the gauge coupling $g$, the choice of gauge group $N$, the VEV $v$, the effective Higgs mass $\mu_\mathrm{eff}$, and the dark matter mass $m_\chi$ supplemented with the initial dark sector temperature $\xi$ and the fraction of $\chi$ dark matter $f_\chi$. These microscopic parameters can be traded for the phenomenological parameters $\Delta N_\mathrm{IR}$, $z_\mathrm{stop}$, $z_*$, $r_g$ and $(\mathcal{G/\mathcal{H}})_\mathrm{ini}$, introduced above. 
To have a first test of the decoupling feature against cosmological data, we  introduce a simplified model in the next section and leave a full exploration of the Hot NEDE phenomenology for future work.

\subsection{Dark radiation matter decoupling model}\label{sec:DRMD}

As mentioned before, the simplified DRMD model arises from the Hot NEDE model in the large-$N$ limit of $SU(N)$ and for a phase transition redshift  $10^6 \lesssim z_* \lesssim  10^9$, in agreement with the bounds in \eqref{z_range}. We recall that here we impose the upper bound to ensure that the phase transition occurs after BBN, thereby satisfying BBN constraints on $\Delta N_\text{eff}$. Modes that are observed in CMB or LSS data are, during these early times, still frozen outside the horizon, and it is sufficient to impose adiabatic initial conditions for the post-phase-transition fluid, effectively eliminating $z_*$ as a free parameter. Moreover, $r_g \to 0$ for $N  \to \infty$, which means that the `second step' in $\Delta N_\text{eff}$ from the mass-threshold effect of the Higgs is negligible. We are therefore left with the additional parameters $\Delta N_\mathrm{IR} = \Delta N_\mathrm{NEDE} \equiv \Delta N_\mathrm{eff} $, $f_\mathrm{idm}$, $z_\mathrm{stop}$ and $(\mathcal{G/\mathcal{H}})_\mathrm{ini}$. 
The DRMD model is thus a four-parameter extension of $\Lambda$CDM. Later, we will see that $(\mathcal{G/\mathcal{H}})_\mathrm{ini}$ is unconstrained by current data sets, making the model effectively a three-parameter extension of $\Lambda$CDM. Here we employ the conventional notation $f_\mathrm{idm} = f_\chi =f_{\chi_0} + f_{\chi_j} $ for denoting the (conserved) fraction of $\chi$ dark matter, and $\delta_\text{idm}\equiv \delta_\chi$ as well as $\theta_\text{idm}\equiv\theta_{\chi}$ for density and velocity perturbations, respectively. As far as CMB and LSS perturbation modes are concerned, the only non-trivial feature that makes this simplified model different from (post-BBN) SIDR is the time-dependent interaction between dark radiation and dark matter. 
In particular, in the limit where $f_\mathrm{idm} \to 0$, both parameters $z_\mathrm{stop}$ and $(\mathcal{G/\mathcal{H}})_\mathrm{ini}$ become unconstrained and the DRMD model reduces to SIDR.\footnote{Later we will see that the same degeneracy leads to strong projection effects when performing a Bayesian parameter extraction for the DRMD model. We therefore perform also a profile likelihood analysis.} In summary, the DRMD and the SIDR model are phenomenological descriptions of the Hot NEDE model for large $z_*$ and $N$, corresponding to the case with and without interacting dark matter, respectively.   

We implement standard adiabatic initial conditions for superhorizon modes with wavenumber $k \tau \ll 1$. In synchronous gauge, they read~\cite{Ma:1995ey}
\begin{align}\label{delta_ini}
\frac{4}{3}\delta_\mathrm{idm} =  \delta_\mathrm{DR}= \delta_\mathrm{\gamma} = \delta_\mathrm{\nu}= -\frac{2}{3} C k^2 \tau^2\,,
\end{align}
where $C$ is an integration constant that can be matched to the (conserved) co-moving curvature perturbation and $\delta_\gamma$ and $\delta_\nu$ are the photon and neutrino density contrast. Solving the coupled perturbation equations for the metric ($h$, $\eta$) and fluid perturbations ($\delta_g$, $\delta_\mathrm{idm}$, $\delta_\nu$, $\delta_\mathrm{DR}$, $\theta_\mathrm{idm}$,  $\theta_\nu$, $\theta_\mathrm{DR}$, $\theta_g$) at leading nontrivial order in $k \tau$, we obtain the following initial conditions for the dark radiation and dark matter velocity perturbations 
\begin{align}\label{theta_dr_ini}
\theta_\mathrm{DR} = \theta_\mathrm{\gamma} = - \frac{1}{18} C k^4 \tau^3 \,,
\end{align}
and
\begin{align}\label{theta_idm_ini}
\theta_\mathrm{idm} = \frac{(\mathcal{G}/\mathcal{H})_\mathrm{ini}}{4 + (\mathcal{G}/\mathcal{H})_\mathrm{ini}} \theta_\mathrm{\gamma}\,.
\end{align}
In particular, we find $\theta_\mathrm{idm} \to 0 $ for $\mathcal{G}/\mathcal{H} \to 0$ as expected for dark matter perturbations in synchronous gauge~\cite{Ma:1995ey} and $\theta_\mathrm{idm} = \theta_\mathrm{DR}$ in the tight-coupling limit  $\mathcal{G}/\mathcal{H} \to \infty$. We note that these initial conditions agree with those used in the related analysis in~\cite{Schoneberg:2023rnx} (and with~\cite{Chacko:2016kgg} in the limit of $\mathcal{G}/\mathcal{H} \to \infty$ and after transforming to synchronous gauge). 
We can also check that for superhorizon modes the adiabatic initial conditions and the perturbation matching in \eqref{eq:matching} become indistinguishable, up to corrections that are suppressed by powers of $k\tau$.

Let us stress that the DRMD model can also be understood as a rather general phenomenological model, which might admit other microscopic realizations. At face value, it describes the decoupling between (self-interacting) dark radiation and a fraction of dark matter through an initially strong interaction that shuts off exponentially. Importantly, here the dark radiation remains fluid-like also after this decoupling (in contrast to photons after decoupling from baryons, and in contrast to atomic dark matter~\cite{Kaplan:2009de,Cyr-Racine:2012tfp,Cyr-Racine:2021oal,Blinov:2021mdk,Bansal:2022qbi}), and $\Delta N_\text{eff}$ is constant from before the onset of the decoupling process until $z=0$ (in contrast to~\cite{Schoneberg:2023rnx}, for which the step in $\Delta N_\text{eff}$ plays a central role in improving the fit to CMB data).

\begin{figure}[t]
    \centering
    \includegraphics[width=\linewidth]{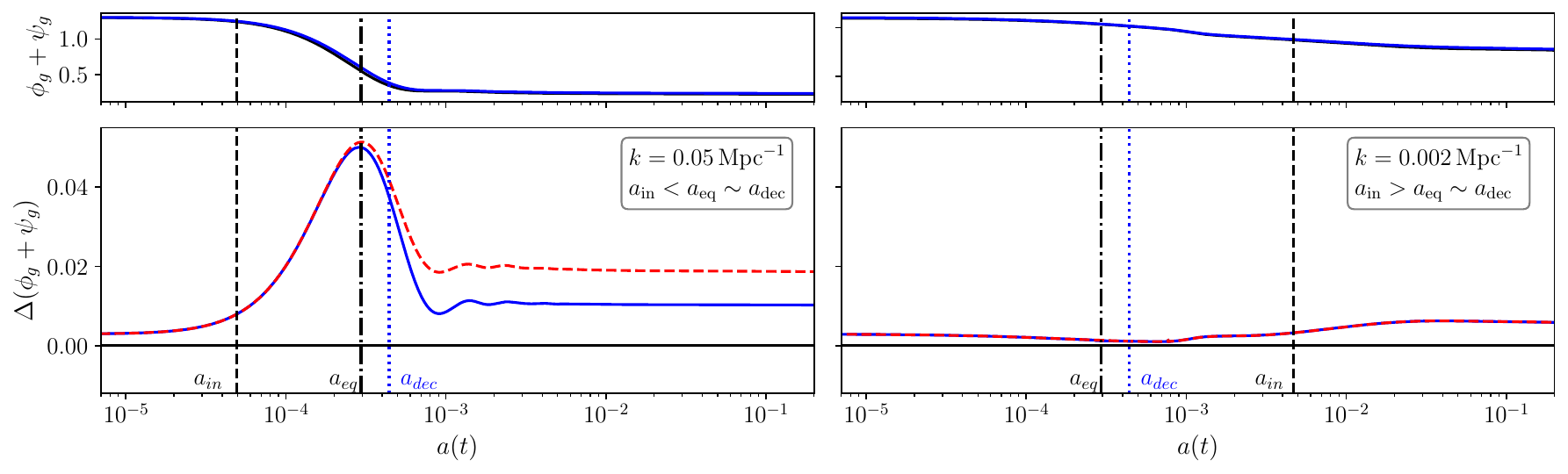}
     \vspace*{-0.5em}

    \includegraphics[width=0.55\linewidth]{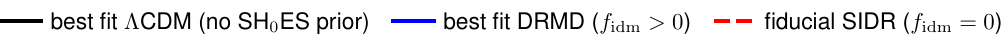}

    \caption{Evolution of the Weyl potential $\phi_g+\psi_g$ for a mode that entered the horizon ($a_\text{in}$) before (left panel) and after (right panel) matter-radiation equality ($a_\text{eq}$). In the lower panel we normalize the amplitudes with respect to the $\Lambda$CDM best fit cosmology (without SH$_0$ES prior). The blue line is the DRMD best fit (with SH$_0$ES prior). For explicit parameter values refer to Tab.~\ref{tab:MCMC}. The red dashed line corresponds to a fiducial SIDR cosmology with the same best fit parameters as the DRMD cosmology except for $f_\mathrm{idm}=0$. In that case, the large value of $H_0$ is penalized by a net enhancement of the gravitational potential for pre-equality modes (left panel). The decoupling feature of the DRMD model compensates this effect while keeping the post-equality mode invariant, provided decoupling ($a_\text{dec}$) occurs around matter-radiation equality. Accordingly, in the right panel, SIDR and DRMD agree.
    }
    \label{fig:Weyl}
\end{figure}

The main effect of the decoupling between dark matter and dark radiation is that modes that entered the horizon after $z=z_\mathrm{dec}$, i.e.\ modes with wavenumber $k\lesssim   \mathcal{H}(z_\mathrm{dec})$, evolve as they would in a standard SIDR scenario without interactions. Modes with $k \gtrsim \mathcal{H}(z_\mathrm{dec})$, on the other hand, `feel' the effects of the interaction. In particular, dark matter perturbations receive a small positive pressure as they are dragged along by dark radiation, which suppresses the gravitational potential and hence the matter power spectrum on those scales. In practical terms, this means that the DRMD model can efficiently compensate for effects that enhance the gravitational potential on small scales. In particular, addressing the Hubble tension through an energy injection in the primordial fluid requires increasing\footnote{This can be understood by a simple geometrical argument: The inverse angular size of the sound horizon (at photon decoupling) $r_s\approx r_d$ is given by $\theta_s^{-1} \simeq \frac{1}{H_0 r_s} \int_0^{z_\mathrm{rec}} dz\, \left[\Omega_m (1+z)^3+(1-\Omega_m)\right]^{-1/2}$. Now, the angular scale $\theta_s$ and $H_0 r_s$ are constrained model-independently at the sub-permille~\cite{Planck:2018vyg} and the sub-percent level~\cite{DESI:2025zgx}, respectively. As we consider models that leave the physics of recombination unchanged, $z_\mathrm{rec}$ is fixed to its $\Lambda$CDM value. This implies that $\Omega_m$ can only deviate from its $\Lambda$CDM value by a few percent, and thus increasing $h$ by around $8\%$ to resolve the Hubble tension is accompanied by a $> 10\%$ increase in $\omega_m = \Omega_m h^2$.}$\omega_m = \Omega_m h^2$ by up to $15 \%$. Such an increase is problematic. For example, it leads to an earlier onset of matter domination, which in turn reduces the net decay of the gravitational potential for any mode that entered the horizon \textit{before} matter domination. This effect can be seen by comparing the left and right panels in Fig.~\ref{fig:Weyl}, which show the evolution of the sum of the gravitational potentials $\psi_g$ and $\phi_g$ in conformal Newtonian gauge (adopting the convention in \cite{Ma:1995ey}) for a pre- and post-phase transition mode, respectively. In both models, SIDR (red dashed) and DRMD (blue), there is an enhancement of the gravitational potential.
As a consequence, the success of these models relies crucially on their ability to counter the effects of an increase in $\omega_m$.  While this is achieved through different mechanisms such as delaying matter domination or having sizeable dark sector acoustic oscillations (see \cite{Lin:2019qug,Niedermann:2020dwg} for more details), the interactions between dark matter and dark radiation in the DRMD model provide an additional mechanism operative on scales $k   \gtrsim \mathcal{H}(z_\mathrm{dec})$. This is visualized in Fig.~\ref{fig:Weyl}, which compares the best fit DRMD cosmology with a fiducial SIDR cosmology (red dashed) with the same parameter values, except that dark radiation in the SIDR model does not interact with dark matter, i.e.\ $f_\mathrm{idm}=0$. It can be seen that in the DRMD model the enhancement of the gravitational potential is (partially) compensated, leading to a smaller amplitude of the gravitational potential for pre-equality modes (left panel) compared to SIDR. 
This mechanism is particularly effective if the decoupling occurs during the early stage of matter domination, because then the modes most affected by the increase in $\omega_m$ are subjected to the effects of the interaction, i.e.\ an excess decay of the gravitational potential. We will see in the next section that this expectation, i.e.\ $z_\mathrm{dec} \sim z_\mathrm{eq}$, is confirmed by our parameter extraction, when testing the model against a canonical set of cosmological data.

%%%%%%%%%%%%%%%%%%%%%%%%%%%%%%%%%%%%%%%%%%%%%%%%%%%%%%%%%%%%
\section{Results} \label{sec:results}
%%%%%%%%%%%%%%%%%%%%%%%%%%%%%%%%%%%%%%%%%%%%%%%%%%%%%%%%%%%%

In this section, we first describe the data sets and analysis strategy, and then present our results. We focus on the DRMD model to describe the dark sector, and compare it to the SIDR and $\Lambda$CDM models.

\subsection{Data analysis} \label{sec:MCMC}

\begin{figure}[t]
    \centering

    \includegraphics[width=0.95\linewidth]{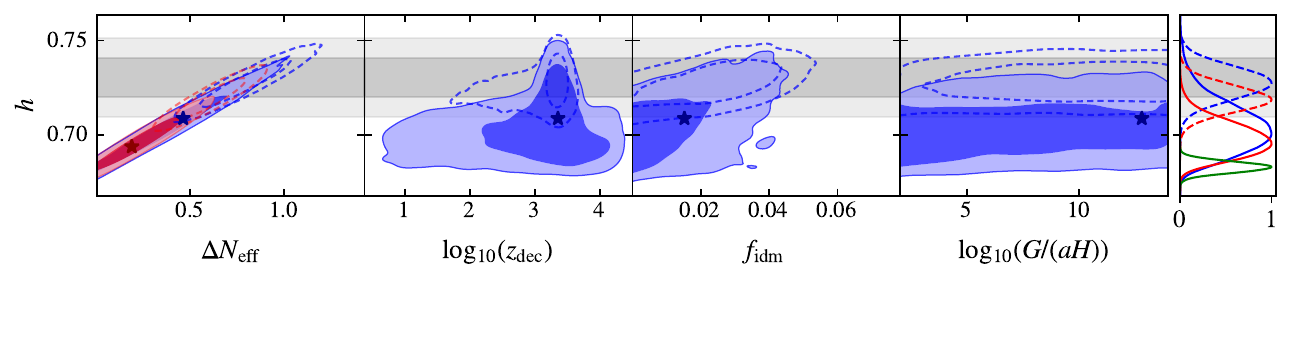}

    \vspace*{-3.5em}

    \includegraphics[width=0.95\linewidth]{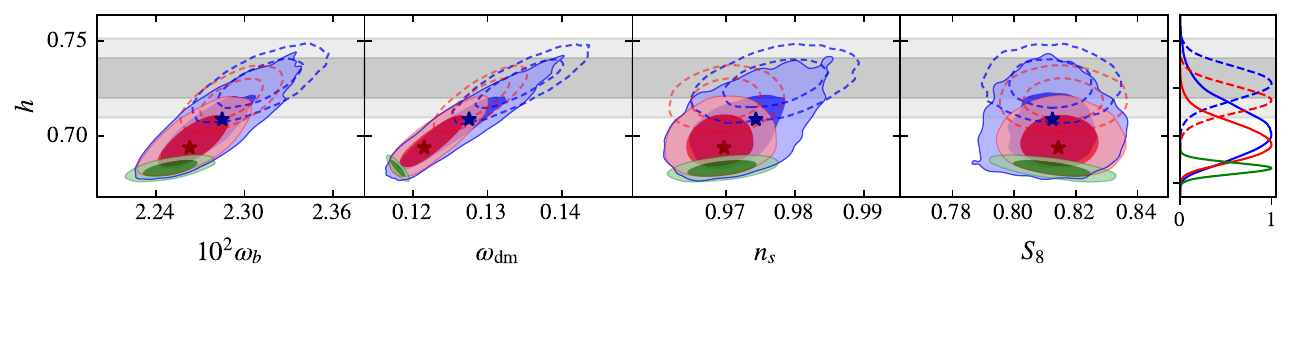}

    \vspace*{-3.5em}

    \includegraphics[width=0.5\linewidth]{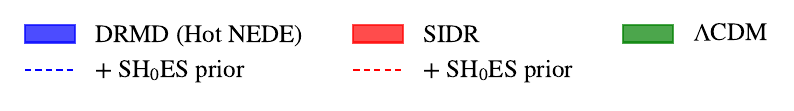}

    \caption{Marginalized posteriors are shown for our baseline data combination, consisting of Planck 2018, DESI BAO (DR2), and (uncalibrated) Pantheon+ SNe data, as filled contours. The dashed contours additionally include a SH$_0$ES prior.  Dark and light shaded regions correspond to $68 \%$ C.L. and $95 \%$ C.L., respectively.  The first and second row display DRMD and $\Lambda$CDM model parameters, respectively. The DRMD model (blue) shows better agreement with the SH$0$ES measurement (gray bands) than SIDR (red) if the decoupling between dark matter and dark radiation occurs at $z_\mathrm{dec} \sim 2000$. Notably, the DRMD contours exhibit strong projection effects when marginalized over the $\log_{10}(z_\mathrm{dec})$ direction, e.g.\ compare the second with the third panel in the first row. As is common for early-time solutions to the Hubble tension, $h$ is strongly correlated with $\omega_b$, $\omega_\mathrm{dm}$, and the spectral tilt $n_s$.  Best fit values of the baseline DRMD and SIDR run, taken from Tab.~\ref{tab:MCMC}, are indicated as blue and red stars, respectively. }
    \label{fig:rectangle}
\end{figure}

We have implemented the  \textit{Dark Radiation Matter Decoupling Model} (DRMD) in the Boltzmann solver \texttt{CLASS} \cite{Blas:2011rf}, dubbed \texttt{DRMD-CLASS}. It is publicly available on GitHub\footnote{\href{https://github.com/NEDE-Cosmo/DRMD-CLASS.git}{https://github.com/NEDE-Cosmo/DRMD-CLASS.git}.},
and includes a component of self-interacting dark radiation (i.e.~described by a fluid) and an additional interacting dark matter component, with background densities characterized by the parameters $\Delta N_\mathrm{eff}$ and the fraction $f_\mathrm{idm}$, respectively. Importantly, the dark radiation component is assumed to be generated after BBN. In practical terms, this means that we take the $\Lambda$CDM value for $N_\mathrm{eff}$ when computing the primordial helium fraction from BBN. 
The interaction between dark radiation and dark matter is parametrized through \eqref{def:G}. It is characterized by two additional model parameters, being the redshift $z_\text{stop}$ at which exponential suppression of the DR-IDM drag rate ${\cal G}$ relative to the Hubble rate ${\cal H}$ sets in, and the initial ratio $({\cal G}/{\cal H})_\text{ini}$  at $z\gtrsim z_\text{stop}$. The decoupling redshift $z_\text{dec}$ is then already fixed by~\eqref{eq:z_ratio2}. For evolving the perturbations, we use the equations in \eqref{eq:delta_DR} and \eqref{eq:delta_idm} with $w_\mathrm{DR} = c_{s}^2=1/3$, in accordance with the assumption that the dark Higgs gives a negligible contribution to dark radiation within the simplified DRMD model. As initial conditions, we use \eqref{delta_ini}, \eqref{theta_dr_ini} and \eqref{theta_idm_ini}. We use the \texttt{DRMD-CLASS} implementation to simulate the $\Lambda$CDM ($\Delta N_\mathrm{eff}$ = $f_\mathrm{idm} = 0$), the SIDR ($f_\mathrm{idm} = 0$), and the DRMD cosmology. The extra parameters compared to $\Lambda$CDM are $\Delta N_\mathrm{eff}$ for SIDR and $\Delta N_\mathrm{eff}, f_\mathrm{idm}, z_\text{stop}, ({\cal G}/{\cal H})_\text{ini}$ for DRMD. For model-selection criteria such as the AIC, we count DRMD as an effective three-parameter extension of $\Lambda$CDM. Although the MCMC analysis samples four additional parameters, the parameters $z_\mathrm{stop}$ and $({\cal G}/{\cal H})_\mathrm{ini}$ are highly degenerate and enter the phenomenology primarily through the derived decoupling redshift $z_\mathrm{dec}$. We therefore apply the AIC penalty to the three phenomenologically constrained directions, while noting that a fully conservative count of all four sampled parameters would shift the quoted DRMD values of $\Delta{\rm AIC}$ upward by $2$.

For our data analysis, we use the following data sets (later referred to as `base'):
\begin{itemize}
\item \bf Planck 2018\rm: CMB anisotropy measurements, which include lensing, high-$\ell$ TT+TE+EE and low-$\ell$ TT+EE data~\cite{Planck:2018vyg,Planck:2018lbu}. 

\item \bf DESI BAO\rm: Release 2 (DR2) BAO data obtained from three years of measurement by the DESI collaboration~\cite{DESI:2025zgx}.\footnote{We use an updated version of the DESI implementation in \href{https://github.com/LauraHerold/MontePython_desilike}{https://github.com/LauraHerold/MontePython\_desilike}.}

\item \bf Pantheon+\rm: SNe data, based on a catalog of 1550 type Ia SNe over the redshift range $0.001 < z < 2.26$~\cite{Brout:2022vxf}.
\end{itemize}

We also carry out complementary runs, where we include the local determination of $H_0$ by the SH$_0$ES collaboration~\cite{Riess:2021jrx} (referred to as `base + SH$_0$ES prior'). We implement this constraint as a Gaussian prior on the SNe type Ia absolute magnitude, which is $M = -19.253 \pm 0.027$ (corresponding to $H_0= 73.04\pm 1.04$ km/s/Mpc).

We perform a Markov Chain Monte Carlo (MCMC) analysis with the \texttt{MontePython} package~\cite{Audren:2012wb,Brinckmann:2018cvx}, employing its implementation of the Metropolis-Hastings algorithm. 
We include as free parameters the six $\Lambda$CDM parameters consisting of the (dimensionless) baryon density $\omega_b = \Omega_b h^2$, the total dark matter density $\omega_\mathrm{dm}=\Omega_\mathrm{dm} h^2$, comprising cold and interacting dark matter, the Hubble parameter $h=H_0/(100 \,\mathrm{km}/\mathrm{sec}/\mathrm{Mpc})$, the primordial scalar power spectrum amplitude $\log_{10}(A_s)$  and tilt $n_s$, and the optical depth to reionization $\tau_{\rm reio}$.  All of them are varied within wide prior ranges. Following the Planck 2018 convention, the neutrino sector is described in terms of two massless and one massive neutrino with mass $m_3=0.06 \,\mathrm{eV}$ and temperature $T_3=0.716 \,T_\gamma$, amounting to $N^{\Lambda\mathrm{CDM}}_\mathrm{eff}=3.046$. 
In the case of the SIDR model, we also sample the additional relativistic density in the form of a (non-free-streaming) fluid component, parametrized in terms of $\Delta N_\mathrm{eff} \in [0,3]$, i.e., the total effective neutrino number is $N_\mathrm{eff}=N^{\Lambda\mathrm{CDM}}_\mathrm{eff}+\Delta N_\mathrm{eff}$. In the case of the DRMD model, these parameters are supplemented with the fraction of interacting dark matter $f_\mathrm{idm} \in [0,1]$, the redshift at which the exponential suppression occurs $\log_{10} (z_\mathrm{stop}) \in [2,5]$, and the initial value of the opacity $\log_{10}\left( \mathcal{G}/\mathcal{H} \right)_\mathrm{ini} \in [2,14]$, where for simplicity we drop the subscript `ini' when presenting results. The prior is chosen such that we always ensure the system to be in a tight-coupling regime when the suppression at $z=z_\mathrm{stop}$ starts. Moreover, $z_\mathrm{dec}$ is inferred from the code as a `derived parameter' using the definition $(\mathcal{G}/\mathcal{H})|_{z=z_\mathrm{dec}} = 1$.

The resulting chains are analyzed with \texttt{GetDist}~\cite{Lewis:2019xzd} and considered converged when the Gelman-Rubin criterion drops below at least $R-1  = 0.05$~\cite{Gelman:1992zz}. For assessing the residual Hubble tension of a given model, we employ the \textit{difference in maximum a posteriori} (DMAP) statistics~\cite{Raveri:2018wln}. It is defined as $Q^{M}_\mathrm{DMAP} = \sqrt{\chi^2(\mathrm{Base}) -\chi^2(\mathrm{Base} + M \,\mathrm{prior})}$ and has the benefit of being insensitive to prior volume effects while reproducing the usual result for Gaussian posteriors. It has become the standard for analyzing Early Universe solutions to the Hubble tension and in particular EDE-type models that are known to exhibit non-Gaussian posteriors~\cite{Herold:2021ksg,Schoneberg:2021qvd}.We also use the Akaike Information Criterion (AIC) $\Delta {\rm AIC} = \Delta \chi^2 + 2\Delta {\rm d.o.f.}$ as a way to penalize extra degrees of freedom compared to $\Lambda$CDM.

Our Bayesian analysis is complemented by a common frequentist approach. It relies on computing the likelihood $\mathcal{L}$ of the data for a given choice of model parameters. To be specific, we employ the \texttt{Procoli} package~\cite{Karwal:2024qpt} to obtain \textit{profile likelihood} curves, which sample the maximal $\mathcal{L}$ along a given axis in parameter space.

\begin{table*}[t]
\renewcommand{\arraystretch}{1.3}
	\centering
		\begin{tabular}{ | l || c | c | c |  c || c | c | c | c ||c|c|}
			\hline
			\multirow{2}{*}{} & 
			\multicolumn{4}{c||}{ { Base} } &
			\multicolumn{4}{c||}{Base + SH$_0$ES prior} & 
			%\multicolumn{3}{|c||}{ + FS + KiDS} & 
            \multirow{2}{*}{{\scriptsize$Q_{\rm DMAP}^{M}$}} &
             \multirow{2}{*}{{\scriptsize$Q_\mathrm{Gauss}$}}\\ 
			\cline{2-9}
			&$H_0$&$\Delta N_{\rm eff}$&$\Delta \chi^2$&$\Delta$AIC&$H_0$&$\Delta N_{\rm eff}$&$\Delta \chi^2$&$\Delta$AIC& &\\ \hline
			
			$\Lambda$CDM  &$68.30 \pm 0.29$& --  & -- & -- &$68.70 \pm 0.29$& --   & -- & --&$5.7 \sigma$&$4.4 \sigma$
			\\ \hline 
			SIDR  &$69.8^{+0.8}_{-1.0}$ ($69.4$)  &$< 0.525$ ($0.188$) & -1.2 & 0.8 &$71.9 \pm 0.7 $ ($72.0$)& $0.61 \pm 0.13$ ($0.610$) & -24.5 & -22.5 &$3.0 \sigma$&$2.5 \sigma$
			\\ \hline 

   			DRMD &$70.5^{+1.0}_{-1.6}$ ($70.9$) &$<0.811$ ($0.462$) & -3.0 & 3.0 &$72.8 \pm 0.8$ ($72.9$)& $0.80 \pm 0.16$ ($0.829$) & -33.3 & -27.3 &$1.4 \sigma$&$1.8 \sigma$ \\

			\hline

		\end{tabular}
	\caption{Comparison between $\Lambda$CDM, (post-BBN) SIDR, and DRMD (Hot NEDE) as solutions to the Hubble tension, using standard metrics. While SIDR reduces the QDMAP tension to $3 \sigma$, the DRMD model brings it to below $2 \sigma$. The (naive) Gaussian tension measure is reported in the last column. After including a prior on $M$, we report a $4.7 \sigma$ and $5 \sigma$ evidence for a non-vanishing fraction of $\Delta N_\mathrm{eff}$ for SIDR and DRMD, respectively. For a complete account of our cosmological parameter inference see Tab.~\ref{tab:MCMC}. 
	}
	\label{tab:resultstension}
\end{table*}

\subsection{Results discussion} \label{sec:resultsdisc}

 As a major result of this work we find that the DRMD model is able to fully resolve the Hubble tension. In particular, we report a small value of DMAP given by $1.4 \sigma$, and hence consistent with statistical fluctuations.\footnote{We emphasize that our analysis always combines BAO, SNe, and CMB data, and is therefore insensitive to possible `internal' tensions among these individual data sets. In particular, the reported $1.4\sigma$ DMAP value refers specifically to the level of agreement between this combined baseline and SH$_0$ES alone. At this stage, we do not investigate how  the recently found~\cite{DESI:2025zgx} $2.3 \sigma$  discrepancy in $\Omega_m$ between DESI DR2 and Planck data is affected within the DRMD and SIDR model.} For $\Lambda$CDM, the DMAP tension is $5.7\sigma$, in agreement with the standard Hubble tension result. For the SIDR model, which we consider for comparing to our results, the DMAP metric yields a significant alleviation, but leaving a residual 3$\sigma$ tension, in broad accordance with previous analyses of non-free-streaming extra radiation that is created after BBN~\cite{Blinov:2020hmc,Schoneberg:2021qvd,Aloni:2021eaq,Schoneberg:2022grr,Joseph:2022jsf,Allali:2023zbi,Khalife:2023qbu,Garny:2024ums}. Thus, the feature of dark radiation matter decoupling within the DRMD model has a decisive impact on the evolution of perturbations, allowing for a significant reduction of the sound horizon, and thus an increase in the inferred $H_0$, while remaining consistent with CMB, BAO and SNe data.
The main results of our MCMC analysis are presented in Tab.~\ref{tab:resultstension}. A more comprehensive table, including mean values, uncertainties, best-fit points, and corresponding $\chi^2$ values, is provided in Appendix~\ref{sec:App_table}. When considering the Gaussian tension, calculated assuming a Gaussian posterior in $H_0$, the tension is reduced from $4.4\sigma$ to $2.5\sigma$ in SIDR and to $1.8\sigma$ in DRMD. We emphasize that, for the baseline data combination without the SH$_0$ES prior, the DRMD model is not preferred over $\Lambda$CDM according to the AIC, yielding $\Delta{\rm AIC}=+3.0$. Thus, without including local $H_0$ information, the baseline data do not provide evidence for the additional DRMD parameters in the model-selection sense, even though they allow substantially larger values of $H_0$ and reduce the tension with SH$_0$ES. On the other hand, when the SH$_0$ES prior is included, the improvement in fit is large enough to overcome the parameter penalty, leading to $\Delta{\rm AIC}=-27.3$.

The DRMD base analysis provides us with the upper bound $\Delta N_\mathrm{eff} < 0.811$ ($95\%$ C.L.), corresponding to an increased mean value for the Hubble parameter of $H_0 = 70.5^{+1.0}_{-1.6}\, \mathrm{km/s/Mpc}$ ($68\%$ C.L.), alongside a fit improvement compared to $\Lambda$CDM of $\Delta \chi^2=-3$ (given the number of free parameters this is at an expected level for the base analysis, i.e.~without including local $H_0$ data; otherwise this becomes $\Delta \chi^2 = -33.3$, see below). This is a clear improvement over SIDR, which leads to the tighter constraint $\Delta N_\mathrm{eff} < 0.525$ ($95\%$ C.L.) and hence a smaller increase in $H_0$. This can also be seen in Fig.~\ref{fig:rectangle} where the filled blue contours corresponding to the DRMD base analysis fully overlap with the SH$_0$ES constraints (within their $95\%$ C.L.), although the latter has not been used in the base analysis. This figure also shows the positive correlations with the $\Lambda$CDM parameters $\omega_b$, $\omega_\mathrm{dm}$, and $n_s$ that are typical for Early Universe solutions to the Hubble tension. However, in contrast to other EDE-type models such as axiEDE~\cite{Poulin:2023lkg} (but notably not Cold NEDE \cite{Cruz:2023lmn}), the value of $S_8$ does not increase. Instead, it is almost unchanged compared to $\Lambda$CDM. We attribute this to the presence of a small fraction $f_\mathrm{idm}< 3.6\%$ ($95\%$ C.L.) of interacting dark matter, which suppresses power on small scales that enter the horizon before $z=z_\mathrm{dec}$.  

It is important to stress that the DRMD analysis suffers from relatively strong projection effects. This can best be seen in the second panel of the first row in Fig.~\ref{fig:rectangle}, where the best fit value, which we indicate as a blue star, lies vertically aligned with the narrow ridge around $\log_{10}(z_\mathrm{dec})= 3.4$ that extends to remarkably high values of $H_0$. This feature would be missed in the other panels that marginalize over  $\log_{10}(z_\mathrm{dec})$ (see also the 1D posterior in the  last panel to the right). We therefore decided to complement our Bayesian analysis with profile likelihood plots, which we present in Fig.~\ref{fig:profile}. Here the left panel corresponds to our baseline (with no prior on $M$). Compared to $\Lambda$CDM, we see that the SIDR model's main effect is to enlarge the width of the 1D profile likelihood, corresponding to an approximate degeneracy between $\Delta N_\mathrm{eff}$ and $H_0$. However, this degeneracy is far from perfect and reaching values of $H_0\sim73 \,\mathrm{km/s/Mpc}$ is still penalized with a $\Delta \chi^2 \sim 10$ (in agreement with a $3\sigma$ residual tension). Moreover, the global best fit value remains close to its $\Lambda$CDM counterpart. This situation changes drastically for the DRMD model. The profile curve becomes very flat, introducing an almost perfect degeneracy between $H_0$ and $(f_\mathrm{idm}, \Delta N_\mathrm{eff})$. At the same time, the global best fit moves to higher values around $H_0=71 \,\mathrm{km/s/Mpc}$. We also indicate the corresponding $1\sigma$ and $2\sigma$ lines, which in agreement with the DMAP tension indicate an insignificant sub-$2\sigma$ residual tension. We stress that this result does not use any local constraints on the expansion rate and is instead driven by CMB, BAO, and (uncalibrated) SNe data alone. 

In the case of the DRMD model, these previous findings justify a combined analysis that includes the SH$_0$ES prior on $M$ (referred to as `Base + SH$_0$ES'). It leads to a relatively high value of the expansion rate of $H_0 = 72.8\pm 0.8\, \mathrm{km/s/Mpc}$ ($68\%$ C.L.), which comes alongside a remarkable $5 \sigma$ evidence for a non-vanishing $\Delta N_\mathrm{eff}$ and a significant fit improvement of $\Delta \chi^2 = -33.3$. The corresponding posteriors are displayed as the dashed contours in Fig.~\ref{fig:rectangle}. While becoming generically tighter and moving to slightly higher values of $H_0$, overall, the DRMD constraints are not changed drastically by including the SH$_0$ES prior (much less than in the case of SIDR, which still exhibits a $3\sigma$ residual tension). In particular, the redshift of dark matter radiation decoupling is only mildly affected and still corresponds to a value close to matter-radiation equality, in agreement with the general discussion in Sec.~\ref{sec:DRMD}. Overall, the late decoupling around $z_\mathrm{dec} \approx 2200 $ (using the best fit in Tab.~\ref{tab:MCMC}) proves to be a rather robust prediction of our model. The relevance of the decoupling mechanism is also reflected by the fact that in the presence of the SH$_0$ES prior the constraint on the interacting dark matter fraction relaxes to $f_\mathrm{idm} < 4.4 \%$ ($95\%$ C.L.), enabling larger values of $H_0$. Correspondingly, Fig.~\ref{fig:profile} shows a degeneracy between $H_0$ and $f_\mathrm{idm}$.

\begin{figure}[!t]
    \centering

    \includegraphics[width=0.99\linewidth]{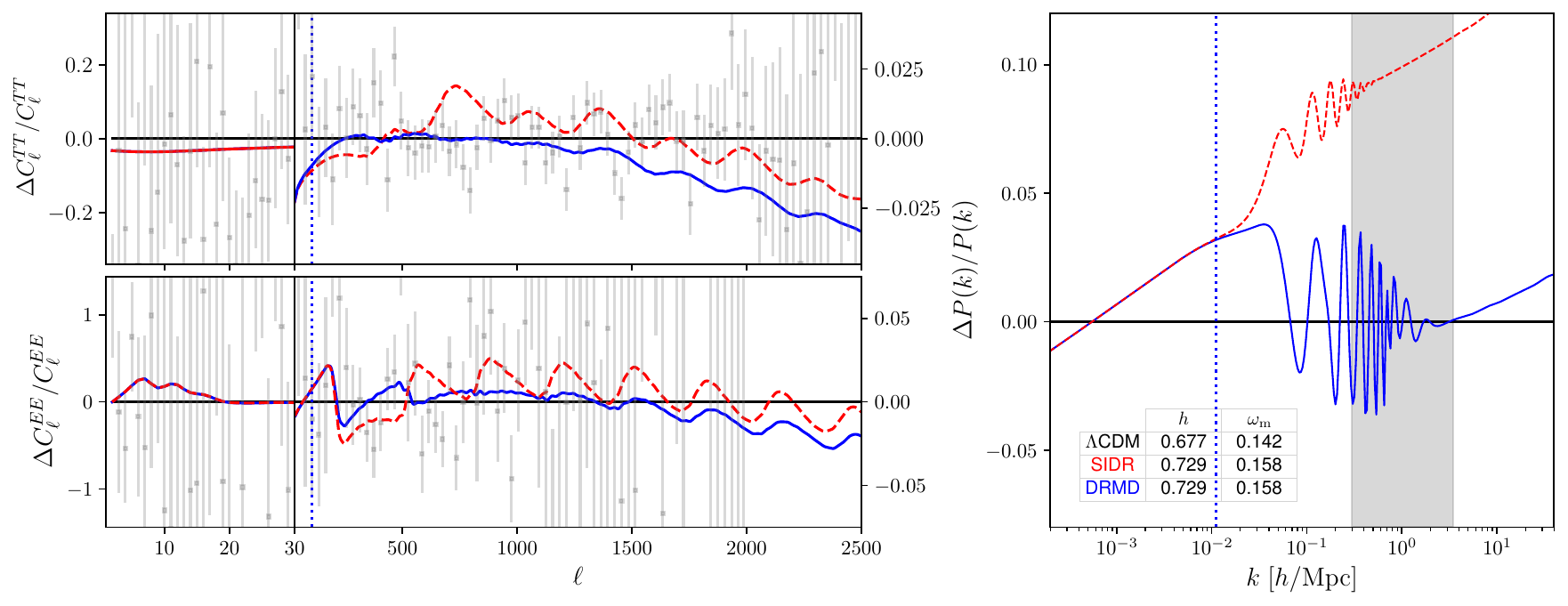}

    \includegraphics[width=0.6\linewidth]{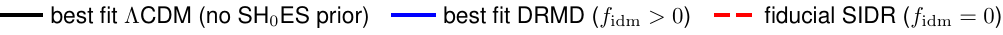}

    \caption{CMB temperature and polarization residuals (left), and the matter power spectrum (right), are shown for the DRMD best-fit cosmology (base with a SH$_0$ES prior), relative to the $\Lambda$CDM base cosmology (base without a SH$_0$ES prior). The red dashed line represents a fiducial SIDR model using the same DRMD best-fit parameters but with $f_\mathrm{idm} = 0$. In the DRMD model, modes to the left of the vertical dotted line have entered the horizon after dark radiation decoupled from dark matter at $z = z_\mathrm{dec}$. This decoupling feature suppresses small-scale power, allowing the model to accommodate large values of $h \simeq 0.73$ without enhancing the matter power spectrum. Parameter values are listed in Tab.~\ref{tab:MCMC}. The two $y$-axes in the left panel refer to the low and high-$\ell$ range separated by the black vertical line.
    }
    \label{fig:spectra}
\end{figure}

On the other hand, the initial coupling strength $\log_{10}(\mathcal{G}/\mathcal{H})_\mathrm{ini}$ is mostly unconstrained. This can be attributed to the fact that for $\mathcal{G}/\mathcal{H} \gg 1$ the dynamics of perturbations becomes universal as $\mathcal{G}$ can be eliminated from the perturbation equations through \eqref{GDelta}. The only way, it still affects the data fit is at a later stage when tight coupling is no longer a valid approximation and $\mathcal{G}/\mathcal{H}$ in \eqref{def:G} approaches values of order unity (around $z \sim z_\mathrm{dec}$), although it is highly degenerate with $z_\mathrm{stop}$. To be specific, by constraining the value of $z_\mathrm{dec}$, due to \eqref{eq:z_ratio}, only the ratio $\ln(\mathcal{G}/\mathcal{H})_\mathrm{ini}/(1+z_\mathrm{stop})$ is determined. This implies we can trade a larger value of $\log_{10}(\mathcal{G}/\mathcal{H})_\mathrm{ini}$ for a larger value of $z_\mathrm{stop}$. In principle, this degeneracy is not perfect, as for example the derivative $d\mathcal{G}/dz|_{z=z_\mathrm{dec}}$ does depend on a different parameter combination. However, as we see in Fig.~\ref{fig:rectangle}, data are almost insensitive to this aspect. This also justifies not discriminating between different decoupling scenarios. As a consequence, we can think about the DRMD model as a three-parameter extension of $\Lambda$CDM that beyond $\Delta N_\mathrm{eff}$ and $f_\mathrm{idm}$ only depends on $(1+z_\mathrm{dec})^{-1}=\ln(\mathcal{G}/\mathcal{H})_\mathrm{ini}/(1+z_\mathrm{stop})$.\footnote{This is also the reason why in Fig.~\ref{fig:rectangle} we choose to present the results in terms of the derived parameter $z_\mathrm{dec}$ rather than $z_\mathrm{stop}$.}  This might however change with future data sets. Let us stress that breaking this degeneracy would allow us to constrain the value of $\alpha_d^2/m_\chi$ through \eqref{eq:z_ratio} and thus directly learn about the range of allowed dark matter masses. Very tentatively, the best fit values in Fig.~\ref{fig:profile} point towards a slight preference for values around $\log_{10}(\mathcal{G}/\mathcal{H})_\mathrm{ini} \approx 12$, which due to \eqref{G_ini_numerical} corresponds to a sub-GeV dark matter mass.

In order to better understand how the DRMD model affects cosmological observables, we show in Fig.~\ref{fig:spectra}  the CMB temperature and polarization residuals (left) alongside the matter power spectrum (right) for the DRMD best fit cosmology (blue line), relative to the base $\Lambda$CDM model (black line). To highlight the relevant features, we consider the DRMD analysis that includes a SH$_0$ES prior. Notably, $\omega_m$ is $12 \%$ larger than in the $\Lambda$CDM reference cosmology. As argued before, this increase enables a large value of $h\simeq 73$ without shifting the CMB peaks. Despite this drastic change in the background cosmology, the power spectra remain comparable to $\Lambda$CDM. In particular, the matter power spectrum does not show a strong enhancement as one could naively expect when $\omega_m$ is increased as much. We  attribute this to the effect of the interactions between dark radiation and dark matter that suppress power on small scales. To illustrate this aspect further, we also included a fiducial SIDR cosmology (red dashed) that uses the same parameter values as the DRMD model but sets $f_\mathrm{idm}=0$. In this case, power is generically enhanced on scales that entered the horizon after the decoupling at $z=z_\mathrm{dec}$ -- represented by the vertical dotted line. Moreover, the fiducial SIDR cosmology exhibits more pronounced oscillatory residuals in the CMB spectra compared to the DRMD model, suggesting that the model struggles to accommodate similarly large values of $H_0$.   
Regarding the matter power spectrum, we note that while the broad-band shape for DRMD remains close to $\Lambda$CDM, the oscillatory part is more pronounced as compared to SIDR  (see right panel of Fig.~\ref{fig:spectra}). This is consistent with the presence of dark acoustic oscillations in DRMD, and represents an important feature that can be directly probed in the nearby future by redshift surveys, as done in the context of BOSS and eBOSS \cite{Beutler:2019ojk,Vasudevan:2019ewf,Mergulhao:2023ukp,Green:2023uyz,Ballardini:2024dto}.

Finally, let us stress that, so far, we have tested the model against a canonical set of cosmological data, including Planck 2018 CMB data. However, from Fig.~\ref{fig:spectra}, we see that the temperature power spectrum for the DRMD (and SIDR) model deviates most significantly from $\Lambda$CDM on small scales, where the Planck error bars become large. As a result, new large-$\ell$ data, as provided recently by ACT~\cite{ACT:2025fju, ACT:2025tim} (and SPT~\cite{SPT-3G:2025bzu}), can test and further constrain our scenario.\footnote{While this was done recently for axiEDE~\cite{Poulin:2025nfb,Khalife:2025xoj}, we stress that the DRMD model, due to its unique decoupling feature, can lead to different conclusions.} We refer the reader to a dedicated follow-up paper~\cite{Garny:2026gcs} for a detailed investigation of SIDR and DRMD in light of ACT data.

To summarize, within the scope of the present analysis, we conclude that the preference for a decoupling feature around matter-radiation equality is driven by Planck data. Whether the simple DRMD model remains compatible with future CMB and LSS data remains to be seen, potentially it will require relaxing assumptions about the primordial spectrum of curvature perturbations.

\section{Conclusions}
\label{sec:conclusion}

We present and analyze a dark sector model described by a simple Lagrangian that is based on renowned principles of particle physics, featuring a non-Abelian gauge symmetry and spontaneous symmetry breaking via a Higgs mechanism. Within our baseline Planck 2018, DESI DR2 and Pantheon+ data combination, we find that its simplified DRMD realization provides a full solution to the Hubble tension. We refer to an implementation capturing the key features as the DRMD model, while the full model is based on the Hot NEDE framework.

From a cosmological point of view, the non-Abelian gauge bosons provide a thermal bath of self-interacting dark radiation. The associated extra relativistic energy density yields a potential mechanism to decrease the sound horizon and thereby enable a larger value of $H_0$ inferred from the position of acoustic peaks in CMB and BAO data, as compared to $\Lambda$CDM. However, as is well known, adding this extra energy without further modifications, known as SIDR model, is by itself not sufficient to resolve the Hubble tension, being disfavored by {\it (i)} its impact on CMB polarization and temperature perturbations as well as {\it (ii)} BBN constraints.

As we show in this work, the feature of spontaneous symmetry breaking within the dark sector simultaneously addresses both of these shortcomings. First, as already noticed in~\cite{Garny:2024ums}, the phase transition associated with symmetry breaking reheats the dark sector, reconciling the cosmic evolution with BBN if the transition occurs between BBN and CMB epochs. Second, adding a fermion multiplet charged under the gauge symmetry yields a component of dark matter that interacts with the thermal bath of gauge bosons. Spontaneous symmetry breaking induces a small mass splitting among the neutral and charged components of the multiplet, implying that this interaction terminates once the dark sector temperature drops below the mass splitting, while the self-interactions within the dark radiation bath remain active due to the non-Abelian self-coupling of gauge bosons. We find that this feature of dark radiation matter decoupling (DRMD) impacts the evolution of perturbations such that CMB polarization and temperature anisotropies as well as BAO are consistent with standard current data sets, while allowing for (and even slightly preferring) a significant dark radiation density and thus an increase in the inferred $H_0$ value.
Specifically, we find $1.4\sigma$ agreement between the combined Planck 2018, DESI DR2, Pantheon+ data sets and SH$_0$ES within a simplified three-parameter implementation of the DRMD model, while they are in tension at $5.7\sigma$ in $\Lambda$CDM and $3.0\sigma$ for SIDR.

From a theoretical standpoint, we stress that the Hot NEDE model provides several non-trivial key features for solving the Hubble tension that follow from a straightforward microscopic model, with well-known dynamics operating at energy scales in the eV-$100$ keV regime close to those of the visible sector, and a Lagrangian given by standard ingredients within particle physics. For comparison, while Cold NEDE
\footnote{The axiEDE model \cite{Poulin:2018cxd} features Planckian axion decay constants and vacuum expectation values within an arguably exotic potential, which is more challenging for embeddings into quantum gravity \cite{Kaloper:2019lpl}, although perhaps not impossible at the cost of some fine-tuning~\cite{Cicoli:2023qri}.} 
is perfectly natural from an effective field theory point of view, featuring very light axion-like particles and GUT scale vacuum expectation values (as is typical for ALP-like models), the involved energy scales are much larger and smaller, respectively, than the energy scale of the NEDE phase transition solving the Hubble tension, and also very different from the energy scales of the visible sector.

Our work motivates several directions for further investigation, such as an analysis of the DRMD model against recent high-resolution CMB data from ACT~\cite{ACT:2025fju} and SPT~\cite{SPT-3G:2025bzu}. Furthermore, in light of future CMB and LSS data, it would be interesting to take into account further features predicted by the model, that are not considered in the simplified DRMD implementation. These can allow for a discrimination against other proposed solutions of the Hubble tension such as Cold NEDE and axiEDE, and include the direct imprint of the dark sector phase transition on perturbations at small scales, leading to features in the matter power spectrum, as well as the mass-threshold effect of the Higgs. Moreover, low-frequency gravitational waves produced during the phase transition provide a potential further signal, which is especially interesting in light of the recent PTA results \cite{EPTA:2023fyk, NANOGrav:2023gor}. From the particle physics perspective, a refined analysis of the decoupling dynamics as well as a study of the early Universe history and the population of the dark sector, for example via (gravitational) freeze-in, are of interest. 
Finally, we stress that the minimal DRMD realization discussed here is strongly constrained by recent ACT high-$\ell$ data when assuming a standard power-law primordial spectrum. As shown in the dedicated follow-up study~\cite{Garny:2026gcs}, however, this conclusion is sensitive to assumptions about the primordial spectrum, and allowing for running and running of the running can restore its viability as a solution to the Hubble tension.

\bigskip

\subsection*{Acknowledgements}

We  acknowledge support by the Excellence Cluster ORIGINS, which is funded by the Deutsche Forschungsgemeinschaft (DFG, German Research Foundation) under Germany’s Excellence Strategy -- EXC-2094 - 390783311. FN was supported by VR Starting Grant 2022-03160 of the Swedish Research Council. We acknowledge Aleksandr Chatrchyan, Lorenzo De Ros and Elisa Ferreira for useful conversations. MSS acknowledges Nordita and MIAPbP for their hospitality and stimulating environments during the completion of parts of this work.

\begin{appendix}

\clearpage
\section{Table with MCMC results}
\label{sec:App_table}
\begin{table}[!th]
	\centering
	\fontsize{8}{9.5}\selectfont
\setlength{\tabcolsep}{4pt}
\begin{tabular}{c|c|c|c|c|c|c}
\hline
\rule{0pt}{3ex}&  \multicolumn{2}{c|}{$\Lambda$\textbf{CDM}}   & \multicolumn{2}{c|}{\textbf{SIDR}}  &  \multicolumn{2}{c}{\textbf{DRMD}}  \\ 
\hline
\hline
\rule{0pt}{4ex}&Base&\makecell{Base\\ + SH$_{0}$ES prior}&Base&\makecell{Base\\ + SH$_{0}$ES prior}&Base&\makecell{Base\\ + SH$_{0}$ES prior}\\
\hline
\rule{0pt}{4ex}\textbf{Parameter}& \makecell{$68 \%$ limits  \\(best fit)}  & \makecell{$68 \%$ limits  \\(best fit)}  & \makecell{$68 \%$ limits  \\(best fit)}  & \makecell{$68 \%$ limits  \\(best fit)} & \makecell{$68 \%$ limits  \\(best fit)} & \makecell{$68 \%$ limits  \\(best fit)} \\ 
\hline
\rule{0pt}{4ex}\makecell{$10^2\omega_b$} & \makecell{$2.250\pm 0.012$\\$(2.243)$} & \makecell{$2.261\pm 0.013$\\$(2.263)$} & \makecell{$2.266\pm 0.016$\\$(2.263)$} & \makecell{$2.291\pm 0.014$\\$(2.292)$} & \makecell{$2.277^{+0.017}_{-0.026}$\\$(2.285)$} & \makecell{$2.310\pm 0.018$\\$(2.313)$} \\
\rule{0pt}{4ex}\makecell{$\omega_{\mathrm{dm}}$} & \makecell{$0.11795\pm 0.00064$\\$(0.11923)$} & \makecell{$0.11710\pm 0.00062$\\$(0.11705)$} & \makecell{$0.1228^{+0.0024}_{-0.0034}$\\$(0.1215)$} & \makecell{$0.1290\pm 0.0026$\\$(0.1291)$} & \makecell{$0.1259^{+0.0034}_{-0.0061}$\\$(0.1275)$} & \makecell{$0.1344\pm 0.0037$\\$(0.1353)$} \\
\rule{0pt}{4ex}\makecell{$h$} & \makecell{$0.6828\pm 0.0029$\\$(0.6771)$} & \makecell{$0.6874\pm 0.0028$\\$(0.6877)$} & \makecell{$0.6978^{+0.0076}_{-0.010}$\\$(0.6937)$} & \makecell{$0.7192\pm 0.0071$\\$(0.7195)$} & \makecell{$0.705^{+0.010}_{-0.016}$\\$(0.709)$} & \makecell{$0.7276\pm 0.0082$\\$(0.7289)$} \\
\rule{0pt}{4ex}\makecell{$\ln(10^{10} A_s)$} & \makecell{$3.054\pm 0.015$\\$(3.048)$} & \makecell{$3.059\pm 0.016$\\$(3.063)$} & \makecell{$3.048\pm 0.015$\\$(3.049)$} & \makecell{$3.042^{+0.014}_{-0.016}$\\$(3.039)$} & \makecell{$3.051\pm 0.015$\\$(3.051)$} & \makecell{$3.050^{+0.015}_{-0.017}$\\$(3.050)$} \\
\rule{0pt}{4ex}\makecell{$n_\mathrm{s}$} & \makecell{$0.9694\pm 0.0033$\\$(0.9671)$} & \makecell{$0.9718\pm 0.0033$\\$(0.9730)$} & \makecell{$0.9692\pm 0.0034$\\$(0.9697)$} & \makecell{$0.9700\pm 0.0034$\\$(0.9697)$} & \makecell{$0.9721^{+0.0039}_{-0.0055}$\\$(0.9743)$} & \makecell{$0.9770\pm 0.0048$\\$(0.9783)$} \\
\rule{0pt}{4ex}\makecell{$\tau_{\mathrm{reio}}$} & \makecell{$0.0611^{+0.0070}_{-0.0078}$\\$(0.0567)$} & \makecell{$0.0639\pm 0.0078$\\$(0.0655)$} & \makecell{$0.0604^{+0.0068}_{-0.0077}$\\$(0.0598)$} & \makecell{$0.0607^{+0.0068}_{-0.0081}$\\$(0.0595)$} & \makecell{$0.0611^{+0.0071}_{-0.0080}$\\$(0.0610)$} & \makecell{$0.0618^{+0.0068}_{-0.0084}$\\$(0.0609)$} \\
\hline
\rule{0pt}{4ex}\makecell{$\Delta N_{\mathrm{eff}}$} & -- & -- & \makecell{$< 0.525$\\$(0.188)$} & \makecell{$0.61\pm 0.13$\\$(0.61)$} & \makecell{$< 0.811$\\$(0.462)$} & \makecell{$0.80\pm 0.16$\\$(0.83)$} \\
\rule{0pt}{4ex}\makecell{$f_{\mathrm{idm}}$} & -- & -- & -- & -- & \makecell{$< 0.0362$\\$(0.0152)$} & \makecell{$< 0.0442$\\$(0.0285)$} \\
\rule{0pt}{4ex}\makecell{$\log_{10}(\mathcal{G}/\mathcal{H})$} & -- & -- & -- & -- & \makecell{unconstrained\\$(12.823)$} & \makecell{unconstrained\\$(13.057)$} \\
\rule{0pt}{4ex}\makecell{$\log_{10}(z_\mathrm{stop})$} & -- & -- & -- & -- & \makecell{$> 2.50$\\$(4.82)$} & \makecell{$> 3.64$\\$(4.83)$} \\
\hline
\rule{0pt}{4ex}\makecell{$\log_{10}(z_\mathrm{dec})$} & -- & -- & -- & -- & \makecell{$3.03^{+0.75}_{-0.34}$\\$(3.36)$} & \makecell{$3.25^{+0.22}_{-0.012}$\\$(3.35)$} \\
\rule{0pt}{4ex}\makecell{$S_8$} & \makecell{$0.8125\pm 0.0084$\\$(0.8242)$} & \makecell{$0.8043\pm 0.0084$\\$(0.8057)$} & \makecell{$0.8151\pm 0.0086$\\$(0.8143)$} & \makecell{$0.8157\pm 0.0085$\\$(0.8146)$} & \makecell{$0.8120\pm 0.0096$\\$(0.8124)$} & \makecell{$0.8120\pm 0.0089$\\$(0.8115)$} \\
\rule{0pt}{4ex}\makecell{$h r_d \mathrm{[Mpc]}$} & \makecell{$100.72\pm 0.48$\\$(99.70)$} & \makecell{$101.46\pm 0.47$\\$(101.51)$} & \makecell{$101.03\pm 0.52$\\$(100.93)$} & \makecell{$101.69\pm 0.47$\\$(101.71)$} & \makecell{$100.95\pm 0.53$\\$(100.96)$} & \makecell{$101.24\pm 0.52$\\$(101.13)$} \\
\rule{0pt}{4ex}\makecell{$\Omega_m$} & \makecell{$0.3026\pm 0.0037$\\$(0.3104)$} & \makecell{$0.2970\pm 0.0035$\\$(0.2967)$} & \makecell{$0.3001\pm 0.0040$\\$(0.3009)$} & \makecell{$0.2950\pm 0.0035$\\$(0.2949)$} & \makecell{$0.3009\pm 0.0040$\\$(0.3007)$} & \makecell{$0.2987\pm 0.0039$\\$(0.2995)$}\\
  \hline
  \hline
  \rule{0pt}{4ex}\makecell{$\mathbf{\chi^2}$ \textbf{per} \\ \textbf{experiment}}&  &   &  &  & &  \\ 
\hline
\rule{0pt}{3ex}Planck high-$\ell$ & 2348.8 & 2351.7 & 2348.9 & 2354.4 & 2347.2 & 2349.5 \\
\rule{0pt}{3ex}Planck low-$\ell$ EE & 397.7 & 399.3 & 397.3 & 397.2 & 397.7 & 397.6 \\
\rule{0pt}{3ex}Planck low-$\ell$ TT & 22.6 & 22.3 & 22.5 & 22.3 & 21.8 & 21.3 \\
\rule{0pt}{3ex}Planck lensing & 9.2 & 9.3 & 9.3 & 9.6 & 9.4 & 9.5 \\
\rule{0pt}{3ex}DESI BAO & 12.8 & 10.4 & 11.8 & 10.4 & 11.6 & 11.0 \\
\rule{0pt}{3ex}Pantheon+ & 1412.7 & 1414.1 & 1413.1 & 1414.5 & 1413.1 & 1413.4 \\
\rule{0pt}{3ex}SH$_0$ES & -- & 29.6 & -- & 3.4 & -- & 0.6 \\
\hline
\rule{0pt}{3ex}\textbf{Total} & \textbf{4203.9} & \textbf{4236.6} & \textbf{4202.9} & \textbf{4211.8} & \textbf{4200.9} & \textbf{4203.0} \\
\rule{0pt}{3ex}$\Delta\chi^2$ & 0.0 & 0.0 & -1.0 & -24.8 & -3.0 & -33.6 \\
\rule{0pt}{3ex}{\scriptsize$Q_{\rm DMAP}^{M}$} [$\sigma$] & -- & 5.7 & -- & 3.0 & -- & 1.4 
\end{tabular}
	\caption{\small Results of our MCMC analysis for $\Lambda$CDM, SIDR and DRMD. The base data sets consists of Planck 2018, DESI BAO (DR2), and Pantheon+. The SH$_0$ES measurement is implemented through a prior on the SNe magnitude $M$. We report the $1\,\sigma$ standard deviation (except for one-sided constraints, which correspond to the $2\,\sigma$ limit). }
    \label{tab:MCMC}
\end{table}

\newpage
%%%%%%%%%%%%%%%%%%%%%%%%%%%%%%%%%%%%%%%%%%%%%%%%%%%%%%%%%%%%%
\section{Dark matter interacting via spontaneously broken \texorpdfstring{$SU(N)$}{SU(N)}}
\label{sec:thermal}
%%%%%%%%%%%%%%%%%%%%%%%%%%%%%%%%%%%%%%%%%%%%%%%%%%%%%%%%%%%%%

In this appendix we provide some details on the mass spectrum and interactions within the dark sector. 

We choose a basis in (dark) $SU(N)$ color space such that the Higgs vacuum expectation value (VEV) is aligned with the $N$th component of the Higgs field $\Psi$,
\be
  \langle\Psi\rangle=(0,\dots,0,v)/\sqrt{2}\,.
\ee

\subsection{Gauge bosons}\label{sec:gaugebosons}

In the unbroken theory, gauge bosons are massless and couple according to the $SU(N)$ generators $t^A$ for $A=1,\dots,N^2-1$,
\be
  D_\mu=\partial_\mu -igt^AA_\mu^A\,.
\ee
Inserting the Higgs VEV into the kinetic term of the Higgs in the Lagrangian yields the mass terms for gauge bosons after symmetry breaking,
\be
  {\cal L}_{\text{gauge boson mass}} = \frac12 g^2 v^2 (t^At^B)_{NN} A_\mu^A A^{B\mu}\,.
\ee
We obtain a mass matrix for the gauge bosons. To analyze it, we consider the generators $t^A$.
The $N^2-1$ generators of the full $SU(N)$ group can be split into three categories: 
\bea\label{eq:generators}
  t^A &=& \underbrace{\left(\begin{array}{cccc} &&& 0 \\ & t^a & & \vdots \\ &&&0 \\ 0&\cdots&0&0
  \end{array}\right)}_{\text{adj of}\ SU(N-1) \atop a=1,\dots,(N-1)^2-1},\quad
  \underbrace{
  \frac12\left(\begin{array}{cccc} 0&\cdots&0&  \\ \vdots & \ddots & \vdots & {\bf e}_j \\ 0&\cdots&0& \\ &{\bf e}_j^T&&0
  \end{array}\right),
  \frac12\left(\begin{array}{cccc} 0&\cdots&0&  \\ \vdots & \ddots & \vdots & i{\bf e}_j \\ 0&\cdots&0& \\ &-i{\bf e}_j^T&&0
  \end{array}\right)
  }_{\text{complex fund. of}\ SU(N-1)\atop j=1,\dots,N-1},\nn\\
  &&
  \underbrace{\frac{1}{\sqrt{2N(N-1)}}\left(\begin{array}{cccc} 1&&& 0 \\ & \ddots & & \vdots \\ &&1&0 \\ 0&\cdots&0&-(N-1)
  \end{array}\right)}_{\text{singlet of}\ SU(N-1)},
\eea
where ${\bf e}_1^T=(1,0,\dots,0),\dots , {\bf e}_{N-1}^T=(0,\dots,0,1)$ are the $N-1$ dimensional unit basis vectors.
The generators are normalized as usual, ${\rm tr}(t^At^B)=\delta^{AB}/2$.

We accordingly split the gauge fields into three categories, 
\be
\begin{array}{cccc}
  \text{adj.} & A_\mu^a & a=1,\dots,(N-1)^2-1 & \text{massless}\,,\\
  \text{fund.} & A_\mu^{\pm j} & j=1,\dots,N-1 &  m_{A^\pm}=gv/2 \,,\\
  \text{sing.} & A_\mu^0 &  &  m_{A^0}=gv \sqrt{\frac{N-1}{2N}}\,. 
\end{array}
\ee
The masses follow from inserting the generators into the mass term. Here we define $A_\mu^{\pm j}\equiv\frac{1}{\sqrt{2}}(A_\mu^{{\bf e}_j}\pm i A_\mu^{i{\bf e}_j})$ where the
gauge fields are those associated to the generators involving ${\bf e}_j$ and $i{\bf e}_j$, respectively. Indeed, the mass term reads
\be
  {\cal L}_{\text{gauge boson mass}} = m_{A^\pm}^2 A_\mu^{+j}A_\mu^{-j} + \frac12 m_{A^0}^2 A_\mu^0A^{0\mu}\,.
\ee
Later we also need some knowledge about triple gauge boson self-interactions. The only non-zero interactions are
of the form adj.$^3$, adj.--fund.--fund., sing.--fund.--fund., see Appendix\,\ref{sec:na}. 

\subsection{Interacting dark matter}

We consider interacting dark matter described by a Dirac fermion multiplet $\chi$, that transforms in the fundamental representation of $SU(N)$,
\be
  {\cal L} \supset \bar\chi( i \slashed{D}-m_\chi)\chi\,,
\ee
with mass $m_\chi\gg gv$, making up a fraction $f_\text{idm}$ of the total cold dark matter abundance.
After spontaneous symmetry breaking, we have to discriminate the components in the $N$-direction,
and the remaining $N-1$ components that transform under the unbroken $SU(N-1)$ subgroup,
\be
 \chi=(\underbrace{\chi_1,\dots,\chi_{N-1}}_{\text{fund.}\ \chi_j},\underbrace{\chi_N}_{\text{sing.},\ \chi_0})\,.
\ee
Loop corrections lead to a mass splitting, making the singlet slightly lighter for the relevant cases with $N>2$ (see Appendix\,\ref{sec:splitting})
\be
  \Delta m_\chi = m_{\chi_j}-m_{\chi_0} = \frac{g^3v}{32\pi}(N-2)\left(1+\sqrt{\frac{2}{N(N-1)}}\right)\,.
\ee
The interactions follow from writing out the covariant derivative in the kinetic term, and inserting the decomposition of gauge bosons.
We get\label{eq:gaugebosoncouplings}
\be
 {\cal L} \supset g\bar\chi t^A\slashed{A}^A\chi = g\bar\chi_i t^a_{ij}A_\mu^a\chi_j+\frac{g}{\sqrt{2}}\left(\bar\chi_0 \slashed{A}^{-j}\chi_j + \bar\chi_j\slashed{A}^{+j}\chi_0\right) 
 + g\left(q_f\bar\chi_j\slashed{A}^0\chi_j+q_0\bar\chi_0\slashed{A}^0\chi_0\right)\,,
\ee
with $q_f=1/\sqrt{2N(N-1)}$ and $q_0=-(N-1)q_f$. From this one can read off the gauge interactions. Note that only the heavier states $\chi_j$ interact with the massless gauge bosons $A_\mu^a$.

\section{Computation of mass splitting}\label{sec:splitting}

Loop corrections lead to a  mass splitting between the first $N-1$ and the last state. This effect is analogous to the Higgsino and Wino
mass splitting, and we can compute it in an analogous way, see e.g.~\cite{Ibe:2012sx}. Here we consider the leading one-loop contribution,
and the limit $m_A \ll m_\chi$.
The one-loop self-energy diagram with an exchange of a gauge boson reads, in Feynman gauge\footnote{In $R_\xi$ gauge, $-ig_{\mu\nu}\mapsto -ig_{\mu\nu}+i(1-\xi)l_\mu l_\nu/(l^2-\xi m_A^2)$.
The additional contribution gives a loop integral 
\be
  \int d^dl \frac{\slashed{l}(\slashed{p}+\slashed{l}+m_\chi)\slashed{l}(1-\xi)}{((p+l)^2-m_\chi^2)(l^2-m_A^2)(l^2-\xi m_A^2)}.
\ee
Using $\{\gamma_\mu,\gamma_\nu\}=2g_{\mu\nu}$, the numerator can be rewritten as 
\be
  \slashed{l}(\slashed{p}+\slashed{l}+m_\chi)\slashed{l}=((p+l)^2-p^2)\slashed{l}-(\slashed{p}-m_\chi)l^2.
\ee
For computing the (on-shell) mass splitting we set $p^2=m_\chi^2$, such that the contribution involving $(p+l)^2-p^2=(p+l)^2-m_\chi^2$ cancels the $\chi$ propagator in the denominator,
and the resulting loop integral vanishes since the integrand is odd under $l\to -l$. The contribution proportional to $\slashed{p}-m_\chi$ also does not give a contribution to the
on-shell mass correction. The naive argument is that it vanishes due to the Dirac equation when acting on an on-shell spinor, $\slashed{p}u_p=m_\chi u_p$. The more precise argument is
that any correction to the self-energy proportional to $\slashed{p}-m_\chi$ yields equal correction to the mass and kinetic terms, see \eqref{eq:SigmaKM} below, and thus their contributions
to the mass shift cancel each other. Thus, the $\xi$-dependent part of the loop does not contribute to the (on-shell) mass, and the results in $R_\xi$ and Feynman gauge agree. This is consistent with gauge invariance of on-shell masses.}
\be
  i\Sigma_{ab}=(ig)^2(t^At^A)_{ab}\int\frac{d^dl}{(2\pi)^d} \gamma^\mu\frac{i}{\slashed{p}+\slashed{l}-m_\chi}\gamma^\nu \frac{-ig_{\mu\nu}}{l^2-m_A^2}\,.
\ee
Here we sum over $A=1,\dots,N^2-1$ gauge bosons, each with mass $m_A$. The indices $a, b=1,\dots,N$ correspond to the color indices of the
in- and outgoing $\chi$ lines, respectively. In the unbroken phase, all $m_A=0$, and the sum over $A$ gives the unit matrix $(t^At^A)_{ab}=C_F\delta_{ab}$
with a constant $C_F$. This means that the self-energy is diagonal in color space in the unbroken phase. Thus the one-loop correction to the $\chi$ mass is
identical for all components $a=1,\dots,N$ before the symmetry breaking. After symmetry breaking, the masses $m_A$ are different for the various gauge bosons $A=1,\dots,N^2-1$.
This leads to a mass splitting. 
Since we are only interested in the splitting relative to each other, it is sufficient to compute (we omit indices $ab$ for brevity)
\be
  \Delta\Sigma = \Sigma - \Sigma\big|_{m_A=0}\,.
\ee
We can generally decompose the self-energy into a correction of the kinetic and of the mass term as
\be\label{eq:SigmaKM}
  \Delta\Sigma = \slashed{p}\Sigma^K(p^2)-m_\chi\Sigma^M(p^2)\,.
\ee
The one-loop mass shift is (see e.g.~\cite{Ibe:2012sx})
\be
  \Delta m_{ab} = (\Sigma^M(m_\chi^2)-\Sigma^K(m_\chi^2)) m_\chi  \,.
\ee
To compute it, we anticipate that the result cannot be divergent, and set $d=4$ except in terms that are divergent in intermediate steps of the computation.
We first rewrite the self-energy as
\be
  i\Sigma_{ab}=-ig^2(t^At^A)_{ab}[(-2\slashed{p}+4m_\chi)I_0-2\gamma_\mu I_1^\mu]\,,
\ee
where
\bea
  I_0 &\equiv& -i\int\frac{d^dl}{(2\pi)^d} \frac{1}{(p+l)^2-m_\chi^2}\frac{}{l^2-m_A^2}\,,\\
  I_1^\mu &\equiv& -i\int\frac{d^dl}{(2\pi)^d} \frac{l^\mu}{(p+l)^2-m_\chi^2}\frac{}{l^2-m_A^2}\,.
\eea
Lorentz invariance implies $I_1^\mu=p^\mu I_1(p^2)$ with a scalar integral $I_1(p^2)=\frac{p_\mu}{p^2} I_1^\mu$.
Thus, at one loop, comparing to \eqref{eq:SigmaKM} gives
\bea
  \Sigma^M(p^2) &=& 4g^2(t^At^A)_{ab}\Delta I_0\nn\\
  \Sigma^K(p^2) &=& 2g^2(t^At^A)_{ab}(\Delta I_0+\Delta I_1)\nn
\eea
where $\Delta I_0=I_0-I_0|_{m_A=0}$ and similar for $I_1$. A standard computation gives 
\be
  I_0 = \frac{\Gamma(2-d/2)}{16\pi^2}\int_0^1dx\, \left[xm_\chi^2+(1-x)m_A^2-x(1-x)p^2\right]^{d/2-2}\,.
\ee
For $d\to 4$ the gamma function has a pole, but also the exponent of the square bracket goes to zero.
Expanding the square bracket around $d=4$ shows that the divergence is independent of the masses and
momenta. Thus it cancels in $\Delta I_0$. Using $(d/2-2)\Gamma(2-d/2)\to 1$ for $d\to 4$ gives
\be
  \Delta I_0 = -\frac{1}{16\pi^2}\int_0^1dx\, \left\{ \ln\left[xm_\chi^2+(1-x)m_A^2-x(1-x)p^2\right]-\ln\left[xm_\chi^2-x(1-x)p^2\right]\right\}\,.
\ee
Setting $p^2=m_\chi^2$ to obtain the mass splitting, and expanding for $m_A\ll m_\chi$, yields
\be
   \Delta I_0 = -\frac{1}{16\pi^2}\frac{\pi m_A}{m_\chi}+{\cal O}\left(\frac{m_A^2}{m_\chi^2}\right)\,.
\ee
One can show similarly that $\Delta I_1$ starts only at order $m_A^2$. Thus at leading order in $m_A/m_\chi$,
\bea
  \Sigma^M(p^2=m_\chi^2) &=& -4\frac{g^2}{16\pi}(t^At^A)_{ab}\frac{m_A}{m_\chi}\nn\\
  \Sigma^K(p^2=m_\chi^2) &=& -2\frac{g^2}{16\pi}(t^At^A)_{ab}\frac{m_A}{m_\chi}\nn
\eea
and therefore
\be
   \Delta m_{ab} = -\frac{g^2}{8\pi}(t^At^A)_{ab} m_A \,.
\ee
This contains a sum over all $A=1,\dots, N^2-1$ dark gauge bosons,
and is a mass-matrix in (dark) color space.

Taking the example of $N=3$, and using the Gell-Mann matrices $t^A=\lambda^A/2$,
as well as the masses $m_{A_1}=m_{A_2}=m_{A_3}=0, m_{A_4}=m_{A_5}=m_{A_6}=m_{A_7}=gv/2$, $m_{A_8}=gv/\sqrt{3}$,
one finds
\be
  (t^At^A)_{ab}m_A=\frac{gv}{4}\left(\begin{array}{ccc}1&&\\&1&\\&&2\end{array}\right) + \frac{gv}{12\sqrt{3}}\left(\begin{array}{ccc}1&&\\&1&\\&&4\end{array}\right)\,.
\ee
Thus the mass matrix is diagonal, and gives an equal mass shift to the $a=1$ and $a=2$ components as required by the remaining $SU(2)$ symmetry after the
phase transition. However, the component $a=3$  receives a more negative mass correction, and is therefore lighter. This is the neutral state under $SU(2)$.
Thus, the splitting between the $a=1,2$ and $a=3$ is
\be
  \Delta m = \Delta m_{11}-\Delta m_{33}= \Delta m_{22}-\Delta m_{33}=\frac{g^3v}{32\pi}\left(1+\frac{1}{\sqrt{3}}\right)\,.
\ee

For general $N$, we use \eqref{eq:generators} and see that the sum of generators corresponding to the fundamental under $SU(N-1)$
gives $\sum t^At^A|_\text{fund} = 1/2\text{diag}(1,\dots,1,N-1)$, and $t^At^A|_\text{sing.}=1/(2N(N-1))\text{diag}(1,\dots,1,(N-1)^2)$.
This means the mass matrix $\Delta m_{ab}$ is diagonal, and splits into a correction to the mass of the first $N-1$ components $\chi_j$ for $a=b=1,\dots,N-1$ (the fundamental
under $SU(N-1)$) and the last component $a=b=N$, $\chi_N\equiv\chi_0$. Altogether,
\bea
  \Delta m_{\chi_j} &=&  -\frac{g^2}{8\pi}\left(\frac12 m_{A^\pm}+\frac{1}{2N(N-1)}m_{A^0}\right)\,,\nn\\
  \Delta m_{\chi_0} &=&  -\frac{g^2}{8\pi}\left(\frac{N-1}2 m_{A^\pm}+\frac{(N-1)^2}{2N(N-1)}m_{A^0}\right)\,.
\eea
This yields for the relative mass splitting
\be
\Delta m = \Delta m_{\chi_j} - \Delta m_{\chi_0} = \frac{g^3v}{32\pi}(N-2)\left(1+\sqrt{\frac{2}{N(N-1)}}\right)\,.
\ee

\section{Computation of conversion cross section}\label{sec:conversion}

Consider $\chi_j\bar\chi_j\to\chi_0\bar\chi_0$ via $t$-channel exchange of $A^{\pm j}_\mu$.
The matrix element for $\chi_i(p_1)\bar\chi_j(p_2)\to\chi_0(q_1)\bar\chi_0(q_2)$ averaged over colors $i,j$ and initial spins,
and summed over final spins reads
\be
  \overline{|{\cal M}^2|} = \frac{g^4}{(N-1)(t-m_{A^\pm}^2)^2}\bigg( 4m_\chi^2(m_\chi+\Delta m)^2-2m_\chi(m_\chi+\Delta m)(p_1q_1+p_2q_2)+2p_1p_2\,q_1q_2+2p_1q_2\,p_2q_1\bigg)\,,
\ee
where $t=(p_1-q_1)^2$. We assume that all in- and outgoing states are non-relativistic and $\Delta m\propto g^3v \ll m_\chi$.
Then we can approximate all summands in the enumerator with $m_\chi^4$, up to corrections that are relatively suppressed by $\Delta m/m_\chi$ or by powers of the velocities,
\be
  \overline{|{\cal M}^2|} = \frac{4g^4m_\chi^4}{(N-1)(t-m_{A^\pm}^2)^2}\,.
\ee
For evaluating the cross section, and the $t$ variable, we introduce a parametrization of the 4-momenta in the center-of-mass frame.
In the center-of-mass frame $p_{1,2}=((m_\chi+\Delta m_\chi)(1+(v_\text{rel}/2)^2/2),\pm(m_\chi+\Delta m_\chi){\bf v}_\text{rel}/2)$
and $q_{1,2}=(m_\chi(1+({v'_\text{rel}}/2)^2/2),\pm m_\chi{\bf v}'_\text{rel}/2)$. Energy conservation fixes the relative velocity in the
final state to be
\be\label{eq:vrelprime}
  {v'_\text{rel}}^2 = v_\text{rel}^2 + \frac{8\Delta m}{m_\chi}\,.
\ee
Here we have neglected terms of order $v_\text{rel}^2\times \Delta m/m_\chi$, being consistent with the assumed power counting.
We are in fact interested in the conversion at temperatures of the dark sector $T_d$ in the ballpark of $\Delta m$.
Assuming that the interacting dark matter is in kinetic equilibrium, this means $m_\chi v_\text{rel}^2\sim T_d$, thus both terms on
the right-hand side of \eqref{eq:vrelprime} need to be kept.
The differential cross section is 
\be
  \frac{d\sigma}{dt} = \frac{1}{64\pi s}\frac{1}{{\bf p}_1^2}\overline{|{\cal M}^2|}\,.
\ee
The boundaries for $t$ are $t_{0,1}=-(|{\bf p}_1|\mp |{\bf q}_1|)^2$. Using $(t_0-t_1)/{\bf p}_1^2=4|{\bf q}_1|/|{\bf p}_1|=4v'_\text{rel}/v_\text{rel}$
and $s=4m_\chi^2$ up to corrections of order $\Delta m/m_\chi$ and $v_\text{rel}^2$ we obtain
\be
  \sigma v_\text{rel} = \frac{v'_\text{rel}}{64\pi m_\chi^2}\frac{4g^4m_\chi^4}{(N-1)}\frac{1}{t_0-t_1}\int_{t_1}^{t_0} dt \frac{1}{(t-m_{A^\pm}^2)^2}
  = \frac{g^4m_\chi^2v'_\text{rel}}{16\pi(N-1)}\frac{1}{(t_0-m_{A^\pm}^2)(t_1-m_{A^\pm}^2)}\,.
\ee
Inserting $t_{0,1}$ and using \eqref{eq:vrelprime} we obtain
\be
  \sigma v_\text{rel} = \frac{\pi\alpha_d^2}{N-1} \frac{m_\chi^2}{(m_{A^\pm}^2+2m_\chi\Delta m)^2+m_\chi^2m_{A^\pm}^2v_\text{rel}^2}\sqrt{v_\text{rel}^2 + \frac{8\Delta m}{m_\chi}}\,.
\ee
We are mostly interested in the conversion cross section when the temperature drops somewhat below the mass splitting, i.e. $T_d\sim m_\chi v_\text{rel}^2 \lesssim \Delta m$.
In this case we can estimate the cross section by its zero-velocity limit
\be
  \sigma v_\text{rel}|_{v_\text{rel}\to 0 } = \frac{\pi\alpha_d^2}{N-1} \frac{m_\chi^2}{(m_{A^\pm}^2+2m_\chi\Delta m)^2}\sqrt{\frac{8\Delta m}{m_\chi}}\,.
\ee

\section{Non-Abelian gauge boson interactions}\label{sec:na}

The generators $t^A$ of $SU(N)$ can be split into three categories as given in \eqref{eq:generators}.
Here we are interested in the self-interaction of gauge bosons after symmetry breaking. We consider
only triple gauge boson interactions, which are sufficient for our purposes. We start
from the full $SU(N)$ case where the triple gauge boson interaction of $A_\mu^A(k), A_\nu^B(q), A_\rho^C(p)$ with all incoming momenta
is given by the Feynman rule
\be
  g f^{ABC}\big( g^{\mu\nu}(k-q)^\rho +g^{\nu\rho}(q-p)^\mu + g^{\rho\mu}(p-k)^\nu \big)\,.
\ee
The interaction is completely determined by the structure constants $f^{ABC}$ of $SU(N)$, defined via
\be
  [t^A,t^B]=if^{ABC}t^C\,.
\ee
We can use the generators from \eqref{eq:generators}. We denote by $a,b,c=1,\dots,(N-1)^2-1$ the indices belonging to the
generators of the unbroken $SU(N-1)$ subgroup. They satisfy the corresponding commutation relations, with structure constants
$f^{abc}$ from $SU(N-1)$. Thus, as expected, the massless gauge bosons feature the usual non-Abelian gauge interactions as dictated
by the residual $SU(N-1)$ gauge symmetry.

Next consider the off-diagonal generators with respect to the $N-1$ subspace and the $N$-component of the broken color direction.
They are given in \eqref{eq:generators}. We denote those involving ${\bf e}_j$ or $i{\bf e}_j$  
by $t^{e_j}$ or $t^{ie_j}$, respectively, for $j=1,\dots,N-1$. It is convenient to define $t^{j\pm}\equiv \frac{1}{\sqrt{2}}(t^{e_j}\mp i t^{ie_j})$.

Finally, the last generator in \eqref{eq:generators} is denoted by $t^0$.
Using the gauge boson notation from Sec.\,\ref{sec:gaugebosons} we have
\be
  A_\mu^At^A = A_\mu^at^a + A_\mu^{j+}t^{j+} + A_\mu^{j-}t^{j-} + A_\mu^0 t^0.
\ee
Using this basis we find that only the following structure constants (plus the ones obtained from arbitrary permutations of the indices) are in general non-zero
\bea
  f^{abc} && \,,\nn\\
  f^{j_\pm 0 j_\pm} &=& \frac{\pm iN}{\sqrt{2N(N-1)}} \,,\nn\\
  f^{j_\pm a i_\pm} &=& \pm i t^a_{ij} \,.
\eea
Note that in the $\pm$ basis, the structure constants are complex (the $\pm i$ prefactors), which is related to the
property that $(t^{j+})^\dag=t^{j-}$.

\end{appendix}
\newpage
\bibliography{ref}
\end{document}